\documentclass[reprint,amsmath,amssymb,aps, superscriptaddress, prx]{revtex4-2} 

\usepackage{amsmath}
\usepackage{amssymb}
\usepackage{amsthm}
\usepackage[cjk]{kotex}
\usepackage{graphicx}
\usepackage{xcolor}
\usepackage{comment}
\usepackage{ulem}
\usepackage{footmisc}
\usepackage{microtype}
\usepackage{hyperref}

\newcommand{\vacuum}{\left | \emptyset \right >}
\renewcommand{\v}{\textbf}

\newtheoremstyle{non-italic} 
  {}
  {} 
  {\normalfont} 
  {} 
  {\bfseries}
  {.} 
  {.5em} 
  {}

\theoremstyle{non-italic}
\newtheorem{theorem}{Theorem}[section] 
\newtheorem{proposition}[theorem]{Proposition}

\newtheorem{lemma}[theorem]{Lemma}

\begin{document}

\title{Itinerant Ferromagnetism from One-Dimensional Mobility}

\author{Kyung-Su Kim (김경수)}
\altaffiliation{Contact author:
\href{mailto:kyungsu@illinois.edu}{kyungsu@illinois.edu}}
\affiliation{Department of Physics and Anthony J. Leggett Institute for Condensed Matter Theory, University of Illinois  Urbana-Champaign, 1110 West Green Street, Urbana, Illinois 61801, USA} 
\author{Veit Elser}
\affiliation{Department of Physics, Cornell University, Ithaca, NY 14853, USA}

\date{\today}

\begin{abstract}
We propose a universal kinetic mechanism for a half-metallic ferromagnet---a metallic state with full spin polarization---arising from strong on-site Coulomb repulsions between particles that exhibit constrained one-dimensional (1D) dynamics.
We illustrate the mechanism in the context of a solvable model on a Lieb lattice in which doped electrons have 1D mobility. 
Such 1D motion is shown to induce only multi-spin ring exchanges of even parity, which mediate ferromagnetism and  result in a unique half-metallic ground state. 
In contrast to the Nagaoka mechanism of ferromagnetism, this result pertains to any doped electron density in the {\it thermodynamic} limit. 
We explore various microscopic routes to such (approximate) 1D dynamics, highlighting two examples: 
doped holes in the strong-coupling limit of the Emery model and vacancies in a two-dimensional Wigner crystal.  
Finally, we demonstrate an intriguing exact equivalence between the bosonic and fermionic versions of these models, which implies a novel mechanism for the conjectured Bose metallic phase.
\end{abstract}
\maketitle

\section{Introduction} 
The half-metallic ferromagnet (or half-metal), a Fermi liquid phase with full spin polarization, is prevalent in nature and represents one of the simplest quantum phases of matter.
Despite its apparent simplicity, understanding its mechanistic origin has long posed a challenge in theoretical physics, as it arises from a subtle interplay between strong interactions and quantum mechanical effects.
The Hubbard model, initially conceived as a model for a half-metal \cite{hubbard1963electron, gutzwiller1963effect, kanamori1963electron, arovas2021hubbard}, was proven to support such a fully polarized ferromagnet \cite{nagaoka1966ferromagnetism} on a {\it finite-sized} bipartite lattice with a {\it single} hole doping in the limit of strong interaction.
However, this result, known as the Nagaoka theorem, does not rigorously establish the existence of a half-metallic {\it phase}, as the thermodynamic limit of the result is singular due to the restriction to single-hole doping. 
More broadly, whether a generic, thermodynamically robust mechanism for a half-metal exists in systems with spin-independent Coulomb interactions has remained an open question.
In this work, we answer this question in the affirmative.

\begin{figure}[b]
    \centering
    \includegraphics[width= \linewidth]{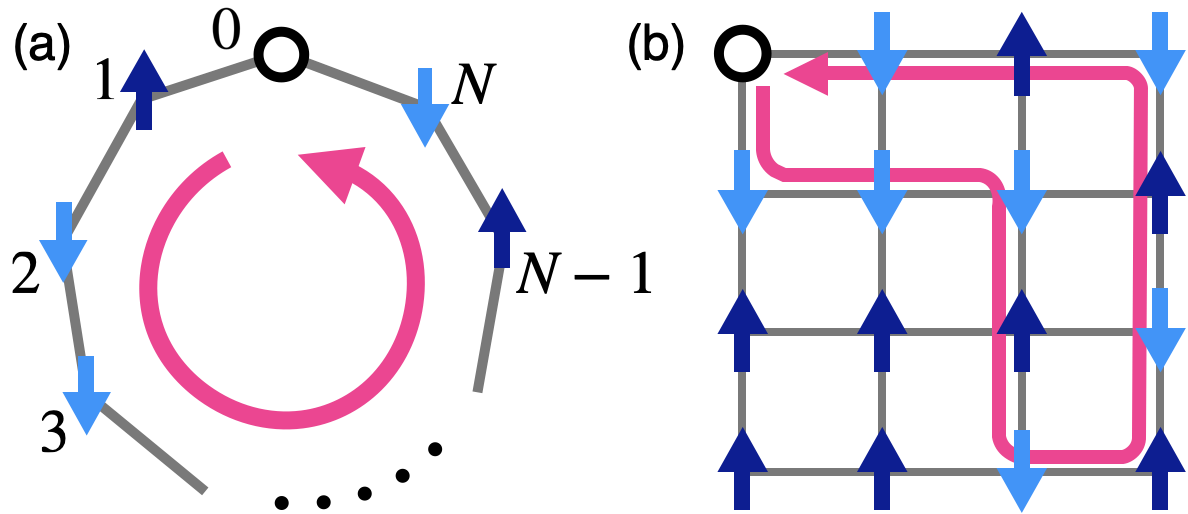}
    \caption{(a) The single-hole problem on an $(N+1)$-site ring \eqref{eq:pinning potential}
    with a pinning potential at site $i=0$.
    The counter-clockwise motion of the hole around the ring induces a multi-spin ring exchange ${\cal P}_N^{-1}$ in a clockwise direction \eqref{eq:N+1 perturbation}.
    (b) On a bipartite lattice, single-hole motion around a loop (shown in magenta) induces a ring exchange involving an odd number of electrons. 
    }
\label{fig:ring_exchange}
\end{figure}

To understand the origin of the exact ferromagnetism in the single-hole case, we invoke the Thouless rule concerning the nature of multi-spin ring-exchange interactions.
To illustrate this rule, let us consider the single-hole problem on a $(N+1)$-site ring in the presence of large pinning potential, $V_0 >0,$ at a site $i=0 \equiv N+1$ [Fig \ref{fig:ring_exchange} (a)]
\footnote{
This Hamiltonian is often written as 
\begin{align*}
    \ \ H_{\rm ring}= P\left [ -t \!\sum_{\sigma=\uparrow,\downarrow}
    \sum_{i=0}^N  \!\left ( c^{\dagger}_{i,\sigma} c_{i+1,\sigma}+ {\rm H.c.}\right )\! + V_0 n_0 \right ] P, 
\end{align*}
where $P$ is the operator that projects out any state with doubly occupied sites.
}
\begin{align}
\label{eq:pinning potential}
    H_{\rm ring}\!=\!  -t \!\sum_{\sigma=\uparrow,\downarrow}
    \sum_{i=0}^N  \!\left ( c^{\dagger}_{i,\sigma} c_{i+1,\sigma}+ {\rm H.c.}\right )\! + V_0 n_0 \!+ [U=\infty].
\end{align}
Here, $n_0 = \sum_{\sigma = \uparrow,\downarrow} c^{\dagger}_{0,\sigma}c_{0,\sigma}$ is the number operator at site $ i = 0$ and the last term prohibits the double occupancy on each site. 
When $t/V_0 = 0$, the hole is pinned at site $i=0$ with all spin states fully degenerate.
This $2^N$ spin degeneracy is lifted for finite $t/V_0>0$; 
the effective magnetic interaction for small $t/V_0$ is governed by $N$-spin ring-exchange interactions induced by the hole motion around the ring, which can be derived at the $(N+1)$th order of perturbation theory
\cite{takahashi1977half}
\begin{align}
\label{eq:N+1 perturbation}
H_{\rm ex} = (-1)^N \frac{t^{N+1}}{V_0^N}\left ( {\cal P}_N + {\cal 
P}_N^{-1} \right ),
\end{align}
where ${\cal P}_N = \sum_{\{\sigma\}}c^{\dagger}_{1,\sigma_{N}}c_{1,\sigma_1} c^{\dagger}_{2,\sigma_{1}}c_{2,\sigma_2} \cdots  c^{\dagger}_{N,\sigma_{N-1}}c_{N,\sigma_N}$ is the spin permutation operator corresponding to the $N$-cycle $\pi_N = (1\ 2\ \cdots\  N).$
Importantly, the sign factor $(-1)^N$ results in the following rule. 
\footnote{Thouless invoked such a rule in trying to understand the nature of magnetic interactions in quantum crystals such as those appear in $^3$He and the two-dimensional electron gas \cite{thouless1965exchange}.
See also Refs. \cite{roger1983RMP, Roger1984WKB, chakravarty1999WC, voelker2001disorder, Katano2000WKB, kim2021discovery, kim2022interstitial, kim2023dynamical}.
}

\smallskip

\noindent
{\bf Thouless rule.} The multi-spin ring-exchange interaction involving an odd (even) number of spins mediates ferromagnetism (antiferromagnetism).
In terms of the parity of permutation, the odd- (even-)spin exchange has even (odd) parity: $(-1)^{N+1} = {\rm sgn} (\pi_N)$.

\smallskip

This insight provides a physical explanation for the exact ferromagnetism in the single-hole problem of the $U=\infty$ Hubbard model on bipartite lattices:
since every loop on a bipartite lattice is of even length, hole motion around any such loop induces an odd-spin exchange and mediates ferromagnetism. 
\footnote{Kinetic antiferromagnetism on non-bipartite lattices, such as those explored in  \cite{haerter2005kineticAF, kim2023exact,kim2024exact}, similarly follows from the Thouless rule.
The hole motion around the smallest triangular loops generates two-spin exchanges that mediate antiferromagnetism.
}
See Fig. \ref{fig:ring_exchange} (b) for an illustration of such a  process on a square lattice.
Notably, the stability of this fully spin-polarized state in the thermodynamic limit has not been  analytically established, except in one dimension, where a perturbative understanding exists \cite{daul1998ferromagnetic}.
Nevertheless, sufficient numerical evidence suggests that such ferromagnetism persists at finite hole densities \cite{emery1990phase, liu2012phases}, up to a critical density beyond which it becomes unstable \cite{shastry1990instability}.

In this work, we propose a novel mechanism for half-metallic ferromagnetism arising from particles with one-dimensional (1D) mobility in generalized $U=\infty$ Hubbard models. 
To illustrate this, we first present a solvable model on a Lieb lattice with doped electrons undergoing  uni-directional, two-particle correlated hopping processes (Sec. \ref{sec:Solvable model for a half-metal}). 
Although the motion of a single electron with 1D mobility cannot induce  magnetism, the correlated motion of multiple such electrons generates magnetic correlations through nontrivial multi-spin ring exchanges.
We show that each of these ring-exchange permutations is of even parity and thus mediates ferromagnetism, resulting in a fully spin-polarized ground state.
The metallic nature of this ground state can be straightforwardly verified, as the Hamiltonian is effectively non-interacting in the ferromagnetic sector.
Combining these results, we establish the half-metallic ferromagnet as a unique ground state for arbitrary doping, apart from a trivial degeneracy due to spin rotational symmetry and gapless Fermi surface excitations.
This starkly contrasts with the Nagaoka mechanism of ferromagnetism, which has only been rigorously established for {\it single} hole doping.

Particles with approximate 1D mobility can arise from strong electron-electron interactions and/or highly anisotropic phonon cloud surrounding a quasi-particle, as illustrated by two examples: 
doped holes in the Emery model in the strong-coupling limit  \cite{kivelson2004quasi1D} (Sec. \ref{sec:Emery model}) and vacancy defects in a two-dimensional Wigner crystal \cite{kim2023dynamical} (Sec. \ref{sec:Wigner crystal}).
While there is competing tendency toward full nematicity (or orbital ferromagnetism) at dilute doping of the 1D particles in these systems, an isotropic half-metallic ferromagnet is expected to be the unique ground state across a range of intermediate doping.
Finally, as a corollary of the evenness of ring-exchange permutations, we demonstrate that systems hosting 1D particles exhibit an exact boson-fermion equivalence, providing a novel mechanism for the Bose Fermi liquid phase (Sec. \ref{sec:Exact Boson-Fermion Correspondence}).

\section{Half-Metallic Ferromagnetism from 1D mobility}\label{sec:Solvable model for a half-metal}

\begin{figure}[b]
    \centering
    \includegraphics[width= \linewidth]{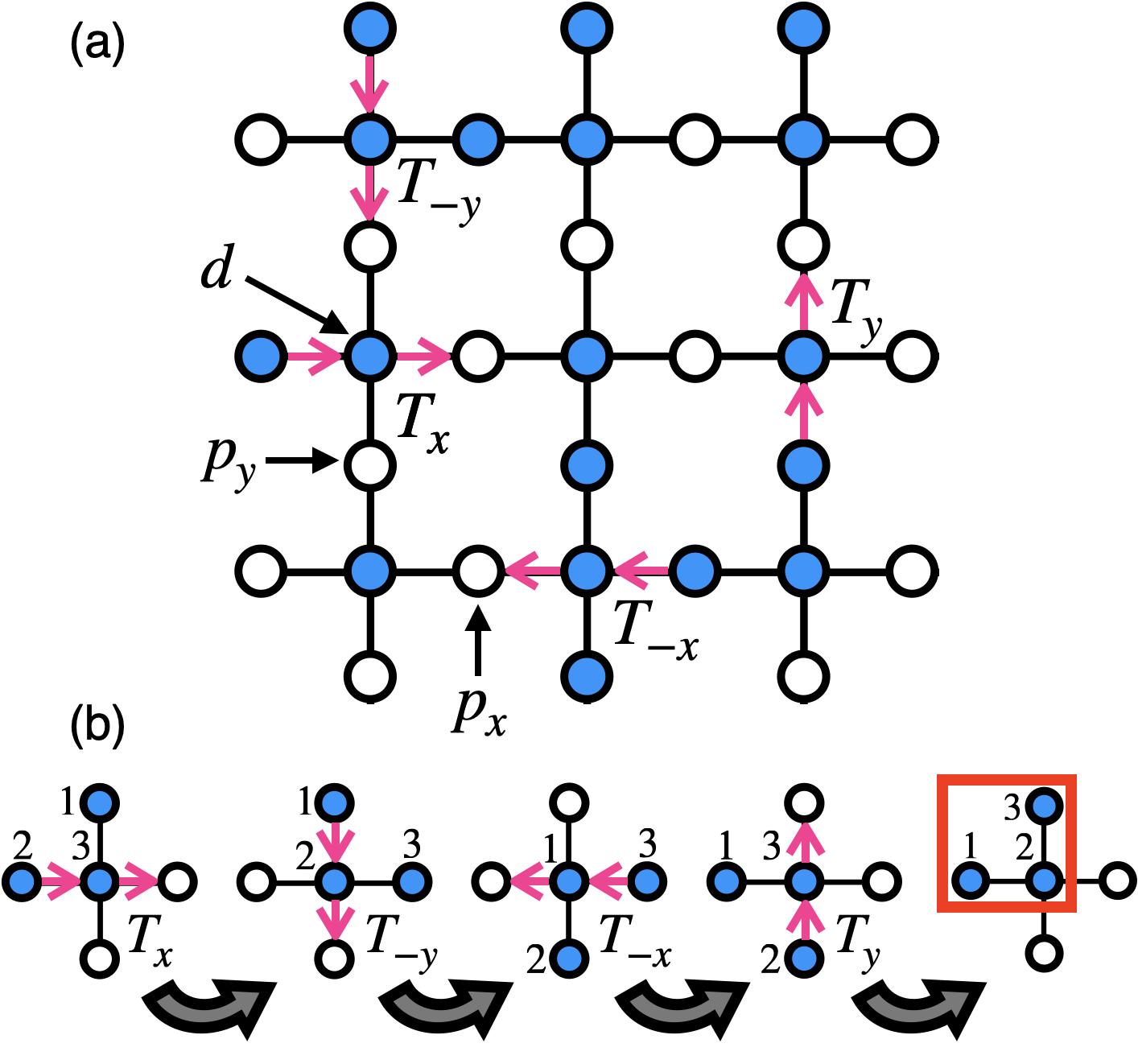}
    \caption{(a) A solvable $U=\infty$  model \eqref{eq:solvable model} on a Lieb lattice under the OBC with dangling bonds on the boundary. 
    $d$ sites are assumed to be
    half-filled (one spinful electron per site), whereas the filling on $p$ sites can vary.
    Here, blue sites are singly occupied and white sites are unoccupied.
    $T_{\pm x}$ ($T_{\pm y}$) are the correlated hopping processes in $\pm x$ ($\pm y$) direction.
    (b) A three-spin ring exchange (or an elementary 3-cycle; red box) induced by a sequence of the 1D hopping processes.
    The site in the middle is a $d$ site and the neighboring sites are $p$ sites.
    }
\label{fig:solvable Lieb lattice}
\end{figure}

To illustrate the concept most clearly, we first consider an exactly solvable model for half-metallic ferromagnetism defined on a Lieb lattice; this model allows the ground state wave function to be explicitly written and analyzed.
The system consists of 
\begin{itemize}
    \item[] 
    \!\!\!\!\!\!\!\!\!\!\!
    {\bf Half-filled $d$ orbitals:} Each $d$-orbital, located at vertices $\v R=i \hat x + j \hat y$ ($i=1,\cdots, L_x$ and $j=1,\cdots, L_y$) of an $L_x \times L_y$ square lattice, is {\it singly} occupied \cite{footnote:filling}. 
    \item[] 
    \!\!\!\!\!\!\!\!\!\!\!
    {\bf Partially filled $p$ orbitals:} The $p$ orbitals are located on the bonds ${\v R + \hat \delta /2}$, where $\delta = x,y$, and their filling may vary.
\end{itemize}
The itinerant $p$-electrons are coupled to the localized $d$-electrons through uni-directional two-electron correlated hopping processes
\begin{align}
\label{eq:solvable model}
    T = -t \sum_{\v R}
    \sum_{\delta = x,y}\sum_{\substack{\sigma,\sigma'  \\ = \uparrow, \downarrow} }
    &\left ( d^{\dagger}_{\v R,\sigma} p^{\dagger }_{\v R + \frac{\hat \delta } 2,\sigma'} d_{\v R,\sigma'}  p_{\v R - \frac{\hat \delta } 2,\sigma}  
    +{\rm H.c.} \right ) \nonumber
     \\
     &\ \ \ \ \ \ \ \ \ \ \ \ \ \ \ \ \ \ \  + [U =\infty],
\end{align}
where the term $[U =\infty]$ enforces a no-double-occupancy constraint on each orbital.
Since $T$ preserves the occupation on each $d$ orbital,  the Hamiltonian dynamics ensures the single-occupancy condition of each $d$ orbital.
As illustrated in Fig. \ref{fig:solvable Lieb lattice} (a), electrons on  $x$ ($y$) bonds are restricted to move along $x$ ($y$) direction.
Such a model with one-dimensional (1D) particles can be motivated from the strong-coupling limit of the Emery model, as we will see later.
We will use both the {\it open boundary condition (OBC)} and the {\it periodic boundary condition (PBC)}.
Under the OBC, we also require that dangling $p$-bonds are present at the boundary [see Fig. \ref{fig:solvable Lieb lattice} (a)].

We intend to establish the existence of a half-metallic  ground state for an entire range of filling of $p$ orbitals, $0<\nu_p<2$, in the {\it thermodynamic limit}, where 
$\nu_p= N_p/N_{\rm vertex}$ is the number of $p$-electrons per unit cell and $N_p$ is the total number of $p$-electrons.
(The total number of electrons then is $N= N_d + N_p,$ where $N_d = N_{\rm vertex} = L_x L_y$ is the number of electrons on the half-filled $d$ orbitals.)
We note that, although we demonstrate our results (Propositions \ref{prop:evenness for Lieb}, \ref{prop:permutation ergodicity OBC}, \ref{prop:permutation ergodicity PBC} and Theorems \ref{thm:exact ferro ground state}, \ref{thm:uniqueness for Lieb}) on a Lieb lattice for concreteness, similar results can be obtained on a larger class of geometries in any dimension with analogously defined 1D hopping processes \eqref{eq:solvable model}.

\smallskip
\noindent
{\bf Subsystem symmetry.} Due to the 1D (or `lineonic') mobility in the model \eqref{eq:solvable model}, the number of $p$-electrons on each row and column is conserved. 
For the case of the PBC, the conserved quantities are:
\begin{align}
\label{eq:conservation_Emery1}
    &X(y) \equiv \sum_{x=1}^{L_x } n_{(x+\frac 1 2,y)},\ \ {\rm for }\ \ y = 1,\cdots,L_y,
    \\
    \label{eq:conservation_Emery2}
    &Y(x) \equiv \sum_{y=1}^{L_y} n_{(x,y+\frac 1 2)} ,\ \ {\rm for }\ \ x = 1,\cdots,L_x,
\end{align}
where $n_{\v r} \equiv \sum_{\sigma}p^{\dagger}_{\v r,\sigma}p_{\v r,\sigma}$.
They are the {\it subsystem symmetries} of the system \cite{footnote:subsystem}.
The conserved quantities can be analogously defined for the OBC.

\smallskip

In the absence of any $p$-electrons, all the $2^{L_xL_y}$ spin states are degenerate.
Such a spin degeneracy is not resolved in the presence of a single $p$-electron under the OBC, due to its 1D mobility that prevents nontrivial exchange among electrons. 
\footnote{
Under the periodic  boundary condition (PBC), the spin degeneracy is partially lifted by the motion of the $p$-electron across the periodic boundary.
This process can be viewed as a boundary contribution to the spin correlations and can be disregarded in the thermodynamic limit.
}
Similarly, the degeneracy is not resolved even in the presence of multiple $p$-electrons of the same bond flavor $\delta$ in the OBC.
Only when there exist at least two $p$-electrons with distinct bond flavors, $\delta = x$ and $y$, is the spin degeneracy lifted through their correlated motions.
An example of correlated moves that result in a non-trivial ring exchange is illustrated in Fig. \ref{fig:solvable Lieb lattice} (b). 
We refer to this three-particle exchange an {\it elementary $3$-cycle}---it will serve as an elementary quantum process that mediates ferromagnetism.

We begin by examining the nature of spin interactions induced by the dynamics of the Hamiltonian \eqref{eq:solvable model}.
In doing so, it is convenient to first treat electrons as distinguishable particles and later impose the antisymmetrization condition: 
positions and spins of these {\it Boltzmannian} electrons are represented by an $N$-tuple (an ordered list), $(\v r_i ) \equiv (\v r_1, \cdots, \v r_N )$ and $(\sigma_i)  \equiv (  \sigma_1, \cdots, \sigma_N )$, respectively, where $N$ is the total number of electrons.
Nontrivial magnetism can arise  only when a particular Boltzmannian-electron configuration, $(\v r_i )$, can be connected to a permuted configuration, $(\v r_{\pi(i)} )$ through repeated applications of terms in the Hamiltonian \eqref{eq:solvable model}, where $\pi \in S_N$ represents some permutation \cite{footnote:distinguishability}.
According to the Thouless rule discussed above, the permutations (or multi-spin ring exchanges) of even parity [${\rm sgn}(\pi) =1$] mediate ferromagnetism, whereas those of odd parity [${\rm sgn}(\pi) =-1$] mediate antiferromagnetism.
The following two results show that any nontrivial ring exchange $\pi$ induced by \eqref{eq:solvable model} is of even parity and thus mediates ferromagnetism.
\footnote{The fact that the evenness of ring-exchange permutations entails ferromagnetism was first recognized by Thouless \cite{thouless1965exchange}. 
This fact can be proven using the path-integral formulation, as shown in Ref. \cite{aizenman1990magnetic}.
}

\begin{proposition} \label{prop:evenness for Lieb}
{\rm \bf{$\pmb{[}$Evenness of Ring Exchange$\pmb{]}$}}

\noindent
Consider the Hamiltonian \eqref{eq:solvable model} on an $L_x \times L_y$ system with either the open (OBC) or periodic boundary condition (PBC).
Additionally for the PBC, we require the following condition:
\begin{itemize}
    \item[(a)] The fermion parity on each row and column must be odd. 
    That is, the number of electrons on each $y$th row and $x$th column, $L_x + X(y)$ and $L_y + Y(x)$, respectively, must be odd. 
\end{itemize}
Then, any permutation $\pi$ of $N$ Boltzmannian electrons induced by the repeated applications of 1D hopping terms in (\ref{eq:solvable model}) is of even parity: ${\rm sgn}(\pi) = 1$.
In other words, $\pi \in {\rm A}_N$, where ${\rm A}_N$ is the alternating group---the group of even permutations---on $N$ objects.
\cite{footnote:Invariant}
\end{proposition}

\begin{theorem}
\label{thm:exact ferro ground state}
{\rm \textbf{$\pmb{[}$Itinerant Ferromagnetism from 1D mobility$\pmb{]}$}}
Under the conditions of Proposition \ref{prop:evenness for Lieb}, the Hamiltonian \eqref{eq:solvable model} hosts the fully spin-polarized state as one of its ground states.
\end{theorem}

{\it Proof of Proposition \ref{prop:evenness for Lieb}.}
We will consider $N_{\rm site} -N$ unoccupied sites to be filled with (Boltzmannian) vacancies.
Specifying the locations of each indexed particle determines a particular (distinguishable) electron-vacancy configuration, which we denote by $N_{\rm site}$-tuple
\begin{align}
\label{eq:R}
    (\pmb{\mathcal{R}}_i)\equiv (\pmb{\mathcal{R}}_1, \cdots ,\pmb{\mathcal{R}}_{N_{\rm site}}).
\end{align}
Here, the indices $i=1,\cdots,N$ are reserved for electrons and $i=N+1,\cdots,N_{\rm site}$ are for vacancies: 
 $\pmb{\mathcal{R}}_i = \v r_i$ for $i=1,\cdots,N$ are initial coordinates of electrons, and $\pmb{\mathcal{R}}_i$ for $i=N+1, \cdots, N_{\rm site}$ are initial coordinates of vacancies.
Then, any other configuration $(\pmb{\mathcal{R}}_i')$ can equivalently be characterized as a permutation of $N_{\rm site}$ objects, $\Pi \in {\rm S}_{N_{\rm site}}$, defined by $\pmb{\mathcal{R}}_i'  = \pmb{\mathcal{R}}_{\Pi(i)}. $
(Note that this is distinct from a permutation $\pi\in {\rm S}_N$ involving only electrons.)
Each term of $T$ \eqref{eq:solvable model} when applied to a particular electron-vacancy configuration, either annihilates that configuration or yields another configuration.
In the latter case, each such term may be interpreted as the action of a 3-cycle on $(\pmb{\mathcal{R}}_i)$ involving two electrons and one vacancy, having an even parity.
Hence, any permutation $\Pi$ obtained from repeated applications of such 3-cycles must also be of even parity  
\begin{align}
\label{eq:invariant for Lieb}
    {\rm sgn}(\Pi)=1.
\end{align}
In other words, the parity of the electron-vacancy permutation is invariant under the dynamics of \eqref{eq:solvable model}.

Now let us restrict ourselves to the final electron-vacancy configuration $(\pmb{\mathcal{R}}_i')$ that has its electron positions  in some permuted locations from their initial positions and similarly for the vacancy positions.
In this case, the associated permutation $\Pi$ involving both electrons and vacancies can be decomposed into the product of permutations involving only electrons ($\pi$) and that involving only vacancies ($\pi_{\rm v}$): $\Pi = \pi \cdot \pi_{\rm v}.$
Then,  $1= {\rm sgn}(\Pi)= {\rm sgn}(\pi){\rm sgn}(\pi_{\rm v}).$ 

In the case of the OBC, it is easy to see that the vacancies in the final configuration $(\pmb{\mathcal{R}}_i')$ cannot be permuted from their initial configuration; 
that is, $\pmb{\mathcal{R}}_i'  =  \pmb{\mathcal{R}}_i$ for $i=N+1,\cdots, N_{\rm site} $  and hence  $\pi_{\rm v} = 1$.
This immediately implies ${\rm sgn}(\pi) = 1$ for the OBC. 
For the PBC, we first note that the fermion parity condition (a) implies that every row or column has an odd number of vacancies, since the total number of sites on each row or  column is even. 
Although the vacancies on the same row or column can cyclically permute, because they are odd in number, such a permutation are of even parity: 
\begin{align}
\label{eq:cyclic permutation of vacancies}
    {\rm sgn}(\pi_{\rm v}) = 1.
\end{align}
Therefore, ${\rm sgn}(\pi) = 1$ for the PBC.
This completes the proof for both boundary conditions.
$\square$

\smallskip

{\it Proof of Theorem \ref{thm:exact ferro ground state}.}
The proof is provided explicitly for the OBC.
The proof for the PBC is essentially the same with minor modifications due to the boundary ring exchange contributions; we will note at specific points, with footnotes, where the changes are needed \cite{footnote:BCforLieb}. 
The proof of the theorem utilizes a choice of many-body basis (referred to as the {\it Perron-Frobenius basis}) that renders the off-diagonal Hamiltonian matrix elements non-positive.
Let $X\equiv \sum_y X(y)$ and  $Y\equiv \sum_x Y(x)$ denote the total number of electrons on $p_x$  and $p_y$ `bonds,' respectively.
The total number of $d$ electrons, again, is given by $N_d = L_x L_y$ and the number of bond sites is $N_{\rm bond } = 2L_xL_y + L_x + L_y$.

We first index the $p$ orbitals in the following way.
Let 
\begin{align}
\label{eq:bond ordering}
    (\v r^{(p)}_i) \equiv 
    (\v r^{(p)}_1,\cdots,\v r^{(p)}_{N_{\rm bond}/2}, \v r^{(p)}_{N_{\rm bond}/2+1},\cdots,\v r^{(p)}_{N_{\rm bond}})
\end{align} 
denote the collection of $p$ orbitals on the $x$- and $y$-bonds. 
Here, the first half of the indices $(1,\cdots , N_{\rm bond}/2)$ correspond to $x$-bonds, ordered first from left to right and then bottom to top, and the last half of the indices  $(N_{\rm bond}/2+1,\cdots , N_{\rm bond})$ correspond to $y$-bonds, ordered from bottom to top and then from left to right. 
For example, under the OBC, $\v r^{(p)}_1,\cdots,  \v r^{(p)}_{L_x+1 }$ index the $x$-bonds on the first row, $\v r^{(p)}_{L_x+2},\cdots,  \v r^{(p)}_{2(L_x+1)}$ index the second row, and so on.
Similarly, $\v r^{(p)}_{N_{\rm bond}/2+ 1},\cdots,  \v r^{(p)}_{N_{\rm bond}/2+L_y+1}$ index the $y$-bonds on the first column, $\v r^{(p)}_{N_{\rm bond}/2+L_y+ 2},\cdots,  \v r^{(p)}_{N_{\rm bond}/2+2(L_y+1)}$ index the $y$-bonds on the second column,
and so on.
Then, one can perform a gauge transformation on $p$ orbitals on every other site
\begin{align}
\label{eq:gauge transformation for solvable model}
    p_{\v r^{(p)}_{i}} \to   (-1)^i p_{\v r^{(p)}_{i}} 
\end{align}
that result in the sign change of the hopping term $t \to -t$. 
The Hamiltonian thus becomes \cite{footnote:T'}
\begin{align}
 \label{eq:solvable'}
   T' = +t \sum_{\v R}
    \sum_{\delta = x,y}\sum_{\substack{\sigma,\sigma'  \\ = \uparrow, \downarrow} }
    &\left ( d^{\dagger}_{\v R,\sigma} p^{\dagger }_{\v R + \frac{\hat \delta } 2,\sigma'} d_{\v R,\sigma'}  p_{\v R - \frac{\hat \delta } 2,\sigma}  
    +{\rm H.c.} \right ) \nonumber
     \\
     &\ \ \ \ \ \ \ \ \ \ \ \ \ \ \ \ \ \ \  + [U =\infty].
\end{align}
The vertex sites $\v 
R_k$ can be ordered in an arbitrary order: $\v R_{1} , \cdots, \v R_{L_x L_y}$.

Electron positions on the $p$ orbitals will be collectively indexed by $I$, where the first $X$ indices correspond to $x$-bonds and the last $Y$ indices correspond to $y$-bonds 
$ I_1 <I_2 < \cdots < I_X \leq  N_{\rm bond}/2 < I_{X+1} < \cdots < I_{X+Y} \leq N_{\rm bond}.$
The spins on $p_x$, $p_y$ and $d$ sites will be collectively denoted by $\sigma \equiv (\sigma_i)$, where
$\sigma_{1}, \cdots, \sigma_{X}$ denote the spins on $x$-bonds $\v r^{(p)}_{I_1},\cdots , \v r^{(p)}_{I_X}$; 
$\sigma_{X+1}, \cdots, \sigma_{X+Y}$ denote the spins on $y$-bonds $\v r^{(y)}_{I_{X+1}},\cdots , \v r^{(x)}_{I_{X+Y}}$; and 
$\sigma_{X+Y+1}, \cdots, \sigma_{N}$ denote the spins on sites $\v R_{1},\cdots , \v R_{N_d}$. 
A basis state is then defined as 
 \begin{align}
\label{eq:basis for Lieb}
    \left | I,\sigma \right >  
     \! 
     \equiv  
     p^{\dagger}_{\v r_{I_1}^{(p)} , \sigma_{1}} 
      \cdots p^{\dagger}_{\v r_{I_{X+Y}}^{(p)} , \sigma_{X+Y}} 
     d^{\dagger}_{\v R_{1}, \sigma_{X+Y+1}} \! \cdots d^{\dagger}_{\v R_{{N_d}}, \sigma_{{N}}} 
     \! \vacuum\!,
\end{align}
where $\vacuum$ denotes a vacuum state with no electrons.

Due to the specific ordering of bond sites, it is straightforward to see that any nonzero matrix element with respect to the basis states \eqref{eq:basis for Lieb} is $-t$, resulting in a non-positive Hamiltonian matrix 
\footnote{
For the PBC, the negativity of the matrix element that moves a $p$-electron across the periodic boundary follows from the assumption that the fermion parity on each row and column is odd, as specified in condition (a) of Proposition \ref{prop:evenness for Lieb}.
See Footnotes \cite{footnote:BCforLieb} and \cite{footnote:T'} for more detail.
}:
\begin{align}
\label{eq:matrix element for $T$}
    \left <I', \sigma'  \right  | T' 
   \left  | I, \sigma    \right  >
 &= 0 \text{ or } -t
 \nonumber \\
 &\equiv 
 -t\  A(I',\sigma';I,\sigma).
\end{align}
Here, we introduced the adjacency matrix $A$ in  Fock space, where the matrix element  $A(I',\sigma';I,\sigma)$ equals 1 whenever the state $\left | I,\sigma \right > $ and $\left | I',\sigma' \right > $ are {\it directly connected} to each other by $T'$ (i.e., $ \left <I',  \sigma'  \right  | T' \left  | I, \sigma    \right  > \neq 0$), and 0 otherwise.
It is also convenient to introduce the adjacency matrix in the ferromagnetic sector
\begin{align}
\label{eq:adjacency in ferro sector for $T$}
    A_{\rm F}(I',I)\equiv A(I',(\uparrow,\cdots,\uparrow);I,(\uparrow,\cdots,\uparrow)).
\end{align}
Note that given an initial basis state $\left | I,\sigma \right >$ and a final orbital configuration $I'$, there is at most one final spin pattern $\sigma'$ that can be directly connected to $\left | I,\sigma \right >$.
We define a function $f$ that gives such a unique spin pattern by 
$\sigma' \equiv f_{I'I}(\sigma)$, when they can indeed be directly connected to each other.
If they cannot be directly connected, $f$ can be defined arbitrarily, for instance, by setting $f =1.$
Then, $A$ can be decomposed as
\begin{align}
\label{eq:decomposition of $A$ for $T$}
A(I',\sigma';I,\sigma) \equiv 
    A_{\rm F}(I',I) \delta(\sigma', f_{I'I} (\sigma)),
\end{align}
where $\delta $ is the Kronecker delta whose value is 1 if $\sigma'$ and $f_{I'I}(\sigma)$ are equal as a tuple and 0 otherwise.

With this setup, we are now in a position to prove the statement of the theorem.
We will do so by explicitly constructing a ferromagnetic state with energy less or equal to that of any given non-ferromagnetic eigenstate.
Let 
\begin{align}
   \left |  \Psi \right >  = \sum_{I,\sigma} \Psi(I,\sigma) \left | I,\sigma
   \right > 
\end{align}
be such an eiegenstate with a unit norm $\left < \Psi|\Psi \right > = 1$, where the sum is taken over basis states within a particular sector of $S_{\rm tot}^z$ (total $S^z$), $X(y)$ and $Y(x)$.
We assume without loss of generality that the coefficients $\Psi(I,\sigma)$ are real, since the Hamiltonian matrix is real \eqref{eq:matrix element for $T$}.
We then construct the following ferromagnetic state
\begin{align}
    \label{eq:trial ferromagnetic state for $T$}
    \left |\Psi_{\rm F} \right >\equiv \sum_{I}\Psi_{\rm F}(I) \left |I, (\uparrow,\cdots,\uparrow) \right >,
\end{align}
where
\begin{align}
        \Psi_{\rm F}(I) \equiv \sqrt{\sum_\sigma [\Psi(I,\sigma)]^2} \geq 0.
\end{align}
$\left < \Psi_{\rm F}|\Psi_{\rm F} \right > = 1$ follows from $\left < \Psi|\Psi \right > = 1$.
That the ferromagnetic state $\left | \Psi_{\rm F} \right >$ has energy less than or equal to that of $\left | \Psi \right >$ can be verified as
\begin{align}
    &\left < \Psi| T' | \Psi \right > = -t \sum_{{I,\sigma, I',\sigma'}} \Psi(I',\sigma') \Psi(I,\sigma) 
    A(I',\sigma';I,\sigma)
    \nonumber \\
    &
    = 
    -t \sum_{{I ,I',\sigma}} \Psi(I', f_{I'I}(\sigma)) \Psi(I,\sigma) 
    A_{\rm F}(I',I)
    \nonumber \\
    &\geq 
    -t \sum_{{I,I'}}
    \sqrt{\sum_{\sigma} [\Psi(I', f_{I'I} (\sigma))]^2} 
    \sqrt {\sum_{\sigma} [\Psi(I, \sigma)]^2} 
    A_{\rm F}(I',I)
    \nonumber \\
    & = 
    \left < \Psi_{\rm F}| T' | \Psi_{\rm F} \right >.
\end{align}
Here, in the third line, we used the Cauchy-Schwarz inequality. 
This concludes the proof of the theorem.
$\square$

\smallskip

While Proposition \ref{prop:evenness for Lieb} and Theorem \ref{thm:exact ferro ground state} establish that the fully spin-polarized state is one of the ground states, they do not guarantee the uniqueness of such a ferromagnetic ground state. 
The following result pertains directly to the uniqueness of the fully spin-polarized ground state across the entire range of filling $0<\nu_p<2$ in the thermodynamic limit.
The proof is technical and will be presented in Appendix \ref{app:proof of ergodicity for Lieb}.

\begin{proposition}
\label{prop:permutation ergodicity OBC}
{\rm\textbf{$\pmb{[}$Even-Permutation Ergodicity; OBC$\pmb{]}$}} 
Consider the Hamiltonian \eqref{eq:solvable model} on an $L_x \times L_y$ Lieb lattice under the open boundary condition (OBC), with dangling bonds at the boundary [see {Fig. \ref{fig:solvable Lieb lattice} (a)}]. 
Consequently, there are $L_x+1$ and $L_y+1$ bond $p$ orbitals on each row and column, respectively. 
For electronic configurations that satisfy the following condition,
\begin{itemize}
    \item[(a)] $0<X(y)<L_x$ and $0<Y(x)<L_y $ for each $y$ and $x$, respectively, meaning that each row and column contains at least one $p$ electron and two vacancies,
\end{itemize}
{\it any even permutation (and none of the odd permutations) of $N$ Boltzmannian electrons can be induced} by repeated applications of terms in \eqref{eq:solvable model}.

In the terminology introduced in the following, the Hamiltonian dynamics is {\it even-permutation-ergodic} whenever condition {(a)} is  satisfied.
\end{proposition}

\noindent
{\bf Permutation and Positional Ergodicity.} Obviously, whether or not this property (italicized) is satisfied depends on the specific electronic configuration. 
If this property holds for a configuration  $(\v r_i)$, we say that the dynamics of  Boltzmannian electrons is {\it even-permutation-ergodic} for that configuration $(\v r_i)$.
This contrasts with the case where the dynamics of a Hamiltonian connects a configuration $(\v r _i)$ to any permuted configuration $(\v r _{\pi(i)})$ for $\pi\in {\rm S}_N$; in such a case we say that the respective dynamics is {\it fully-permutation-ergodic} for $(\v r _i)$. 
If none of the nontrivial permutations can be induced by the Hamiltonian dynamics for $(\v r _i)$, then the specified dynamics is said to be {\it fully-permutation-nonergodic} for $(\v r _i)$.
For example, it is straightforward to see that the single-hole problem in the $U=\infty$ Hubbard model on a square lattice (with an open boundary condition) is even-permutation-ergodic for any hole position, whereas the same problem on a triangular lattice is fully-permutation-ergodic.
\footnote{
The question of whether or not the {even-permutation ergodicity} or {full permutation ergodicity} is satisfied for the single-hole problem in the $U=\infty$ Hubbard model on a given graph is equivalent to the `graph puzzle' (a generalization of the 15-puzzle) problem on the same graph. This problem was completely characterized by Wilson \cite{wilson1974graph}.
}
On the other hand, the $U=\infty$ Hubbard model in one dimension under the OBC is fully-permutation-nonergodic for any hole concentrations.

Furthermore, if the dynamics of a Hamiltonian connects an electronic configuration 
\begin{align}
    \{\v r_i\}\equiv \{\v r_1,\cdots , \v r _N\}
\end{align}
(represented as a set, treating particles as {\it indistinguishable} rather than as a tuple for distinguishable particles) to any other configuration $\{ \v r '_i \} $ consistent with the subsystem symmetries (\ref{eq:conservation_Emery1}, \ref{eq:conservation_Emery2}), we say that {\it positional ergodicity} is satisfied for the electronic configuration $\{\v r_i\}$ under the Hamiltonian dynamics.
For example, if an electronic configuration  $\{\v r_i \}$ is fully nematic, meaning that $p$-electrons occupy only one bond flavor $\delta $, the dynamics specified by \eqref{eq:solvable model} is positionally ergodic within the symmetry sector that contains $\{\v r_i \}$, but is fully-permutation-nonergodic.

\smallskip

\noindent
{\bf Spin Ergodicity.} Both  even-permutation ergodicity and full permutation ergodicity imply  ergodicity in the spin sector in the following sense:
any spin configuration 
\begin{align}
     \left |\{\v r_i, \sigma_{\pi (i)} \} \right > \equiv c^{\dagger}_{\v r_1, \sigma_{\pi (1)}} \cdots c^{\dagger}_{\v r_N , \sigma_{\pi (N)}} \vacuum
\end{align}
that differs from the initial spin configuration 
\begin{align}
    \left |\{\v r_i, \sigma_i \} \right > \equiv c^{\dagger}_{\v r_1, \sigma_1} \cdots c^{\dagger}_{\v r_N , \sigma_N} \vacuum,
\end{align}
by a permutation of spins is connected to the initial configuration by the Hamiltonian dynamics.
This can be simply seen as follows. 
If the Hamiltonian dynamics is even-permutation ergodic for the configuration $(\v r _i)$, any 3-cycle can be induced (assuming there are more than 3 particles).
In the maximum or minimum $S^z_{\rm tot}$ sectors, no nontrivial spin permutation is possible, so there is nothing to prove.
In other fixed $S^z_{\rm tot}$ sectors, in order to transpose two electrons with opposite spins (say $\sigma_1$ and $\sigma_2$), pick another arbitrary electron (say $\sigma_3$) and apply a 3-cycle $\pi_3$ involving these three electrons.
Either $\pi_3$  or $\pi_3^{-1}$ will induce a two-spin exchange between $\sigma_1$ and $\sigma_2$.
This shows that any two-spin exchange can be induced.
Since any multi-spin exchange can always be decomposed into a product of transpositions (or two-spin exchanges), this completes the proof of ergodicity in the spin sector.

\smallskip

It is clear that the positional ergodicity is satisfied under the Hamiltonian dynamics \eqref{eq:solvable model} for any electronic configuration. 
This follows from the fact that,  on each row (or column), any rearrangement of $p$-electrons that preserves $X(y)$ [respectively, $Y(x)$] can be achieved under the specified dynamics. 

The following proposition is an analogue of Proposition \ref{prop:permutation ergodicity OBC} for the PBC.
The proof is provided in Appendix \ref{app:proof of ergodicity for Lieb}.

\begin{proposition}
\label{prop:permutation ergodicity PBC}
{\rm\textbf{$\pmb{[}$Even-Permutation Ergodicity; PBC$\pmb{]}$}} 
Consider the Hamiltonian \eqref{eq:solvable model} with the periodic boundary condition (PBC) on an $L_x \times L_y$ Lieb lattice.
We impose the same condition as in Proposition \ref{prop:evenness for Lieb}:
\begin{itemize}
    \item[(a)] The number of electrons on each $y$th row and $x$th column, $L_x + X(y)$ and $L_y + Y(x)$, respectively, must be odd.
\end{itemize}
Additionally, if the number of electrons occupying $p$ orbitals satisfy the following condition:
\begin{itemize}
    \item[(b)]  $0<X(y)<L_x$ and $0<Y(x)<L_y $ for each $y$ and $x$, respectively, meaning that each row and column contains at least one electron and one vacancy,
\end{itemize}
then the Hamiltonian dynamics is even-permutation-ergodic.
\end{proposition}

The following theorem demonstrates that the even-permutation ergodicity [Propositions \ref{prop:permutation ergodicity OBC} and  \ref{prop:permutation ergodicity PBC}] is sufficient to ensure that a fully spin-polarized metallic ground state is the unique ground state across the entire range of filling $0<\nu_p < 2$ in the {\it thermodynamic limit}, aside from degeneracies arising from spin rotational symmetry and gapless Fermi surface excitations.

\begin{theorem}
\label{thm:uniqueness for Lieb}
$\pmb{[}${\rm \textbf{Unique Half-Metallic Ground State from 1D mobility}}$\pmb{]}$
Under the conditions of  Proposition \ref{prop:permutation ergodicity OBC}, the Hamiltonian \eqref{eq:solvable model} on an $L \times L$ Lieb lattice with the open boundary condition (OBC) hosts a half-metallic ferromagnet as its unique ground state, aside from 
\begin{itemize}
    \item[(i)]  $(N+1)$-fold spin degeneracy associated with spin rotational symmetry, where $N$ is the total number of electrons, and
    \item[(ii)] charge degeneracy associated with $2L$-fold degenerate single particle states on the Fermi surface.
\end{itemize}
\end{theorem}

A similar statement holds for the $L \times L$ system  with the PBC for electronic configurations satisfying the conditions (a) and (b) of Proposition \ref{prop:permutation ergodicity PBC}.
The proof follows essentially the same steps as for the OBC and is therefore omitted here.

\smallskip

{\it Proof of Theorem \ref{thm:uniqueness for Lieb}}.
The proof relies on the Perron-Frobenius theorem for an irreducible non-positive matrix. 
A Hermitian matrix $M$ is said to be irreducible if its associated graph $G_M$---defined to have an edge between $i$ and $j$ precisely when $M_{ij} \neq 0$---is connected.
Eq. \eqref{eq:matrix element for $T$} already demonstrates that the Hamiltonian matrix is non-positive with respect to the basis \eqref{eq:basis for Lieb}.
Furthermore, since positional ergodicity and spin ergodicity hold for the configurations satisfying condition (a), any basis states \eqref{eq:basis for Lieb} with fixed $S^z_{\rm tot }$, $X(y)$ and $Y(x)$ are connected to each other by the Hamiltonian dynamics, establishing the irreducibility of the Hamiltonian matrix.
The Perron-Frobenius theorem then implies that the fully spin-polarized ground state identified in Theorem \ref{thm:exact ferro ground state} is the unique ground state in each symmetry sector specified by $S^z_{\rm tot }$, $X(y)$ and $Y(x)$.
Due to the spin rotational symmetry, the fully spin-polarized state with a maximum $S ^2_{\rm tot}$ and $S^z_{\rm tot}$ is degenerate with states that are obtained from this state by repeated applications of $S^-_{\rm tot}$, yielding the $(N+1)$-fold spin degeneracy noted in (i).

\begin{figure}[t]
    \centering
    \includegraphics[width=0.45\linewidth]{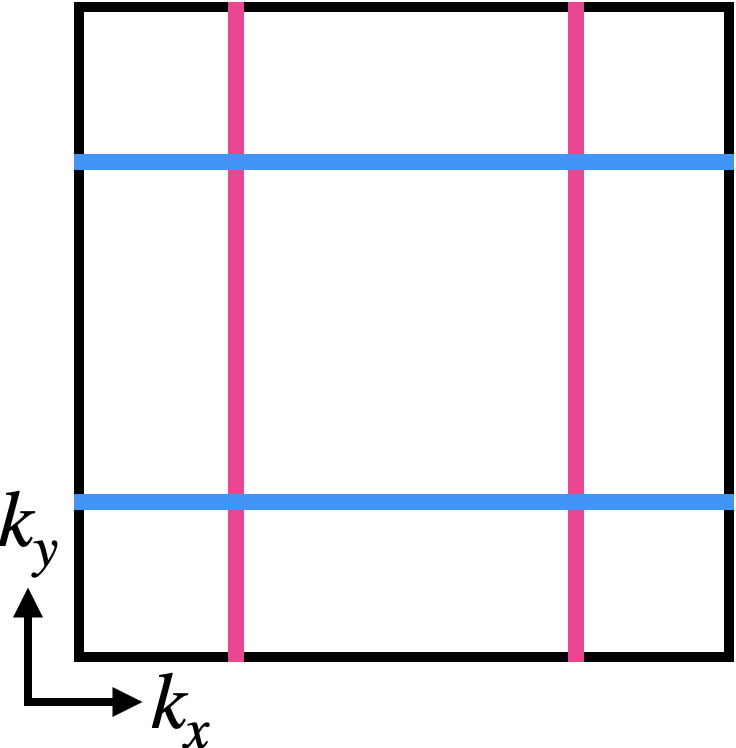}
    \caption{Fermi surfaces of the half-metallic ground state of the model \eqref{eq:solvable model}.
    The black boundaries denote the Brillouin zone; the region interior to the magenta (blue) lines are occupied by $p_x$ ($p_y$) electrons. 
    The bosonic version of the model also exhibits the same Fermi surfaces, as demonstrated  in Sec. \ref{sec:Exact Boson-Fermion Correspondence}
    }
    \label{fig:fermi surface}
\end{figure}

It remains to identify the sector of $X(y)$ and $Y(x)$ that contains the minimum-energy ground state for a given number of electrons $N.$
This is straightforward since in the fully spin-polarized sector (say, when all the spins are pointing in $+z$ direction, $\uparrow$), the Hamiltonian $T'$ \eqref{eq:solvable'} reduces to a non-interacting problem on a $2L$ decoupled wires
\begin{align}
 \label{eq:solvable model in ferro sector}
   T'\vert_{\rm ferro} = -t \sum_{\v R}
    \sum_{\delta = x,y} 
    \left (  p^{\dagger}_{\v R + \frac{\hat \delta } 2,\uparrow}   p_{\v R - \frac{\hat \delta } 2,\uparrow}  
    + {\rm H.c.} \right ).
\end{align}
Each wire has a single particle spectrum 
\begin{align}
\label{eq:band structure}
    E_k = -2t \cos k,
\end{align}
where $k=\frac{\pi n }{L+1} $ ($n=1,2,\cdots, L$).
Since this is a monotonically increasing function of $n,$ the minimum energy configuration is achieved by filling each wire uniformly in their lowest single-particle states.
That is, for $N_p = N - N_d \equiv  2 L d + r$, where $d$ and $r$ are integers satisfying $0 \leq r \leq 2L-1$, single-particle states $n=1,2,\cdots,d$ are filled for each wire, with the remaining $r$ electrons occupying the $n=d+1$ state in any of $\binom{2L}{r}$ ways.
This corresponds to the gapless Fermi surface excitations described in (ii).

The resulting half-metallic ground state has straight line Fermi surfaces, as illustrated in Fig. \ref{fig:fermi surface}.
$\square$

\smallskip

One can consider additional interactions to the Hamiltonian \eqref{eq:solvable model}, such as on-site potentials and density-density interactions,
\begin{align}
    \sum_{\v r} \epsilon_{\v r} n_{\v r}
    + 
    \sum_{\v r,\v r'} V_{\v r,\v r'} n_{\v r} n_{\v r'} +\cdots,
\end{align}
without altering the conclusions of Proposition \ref{prop:evenness for Lieb} and Theorem \ref{thm:exact ferro ground state}.
However, as we will see, the details of interactions can influence whether the ground state belongs to a sector where even-permutation ergodicity and spin ergodicity are satisfied.

Of particular relevance is the term
\begin{align}
\label{eq:NNN interactions}
    V_{pp} \sum_{\v R} &\bigg ( n_{\v R + {\hat x} /2} n_{\v R +{\hat y}/ 2}  +
    n_{\v R - {\hat x}/ 2} n_{\v R +  {\hat y}/ 2}
    \nonumber \\
    &\ \ \ \ \ +
    n_{\v R -  {\hat x}/ 2} n_{\v R -  {\hat y} / 2} +
    n_{\v R +{\hat x}/ 2} n_{\v R -  {\hat y} / 2}
    \bigg ),
\end{align}
which will be shown to induce a fully nematic ground state in the dilute ($0<\nu_p\ll 1$) or dense ($0<2-\nu_p \ll 1 $) electron doping, with electrons or holes occupying only one bond flavor $\delta$, respectively.
In such a case, since no nontrivial permutation can be induced (i.e., the dynamics for the fully nematic configuration is fully-permutation-nonergodic), all spin states are degenerate.

Finally, we note that although we demonstrated the results (Propositions \ref{prop:evenness for Lieb}, \ref{prop:permutation ergodicity OBC}, \ref{prop:permutation ergodicity PBC} and Theorems \ref{thm:exact ferro ground state}, \ref{thm:uniqueness for Lieb}) on a Lieb lattice for concreteness, these results also hold for similarly structured geometries in two or higher dimensions, such as a 2D triangular lattice or a 3D cubic lattice, both with additional sites on each bond.

\section{Quasi-1D hole in Emery model}
\label{sec:Emery model}

\begin{figure}[t]
    \centering
    \includegraphics[width=\linewidth]{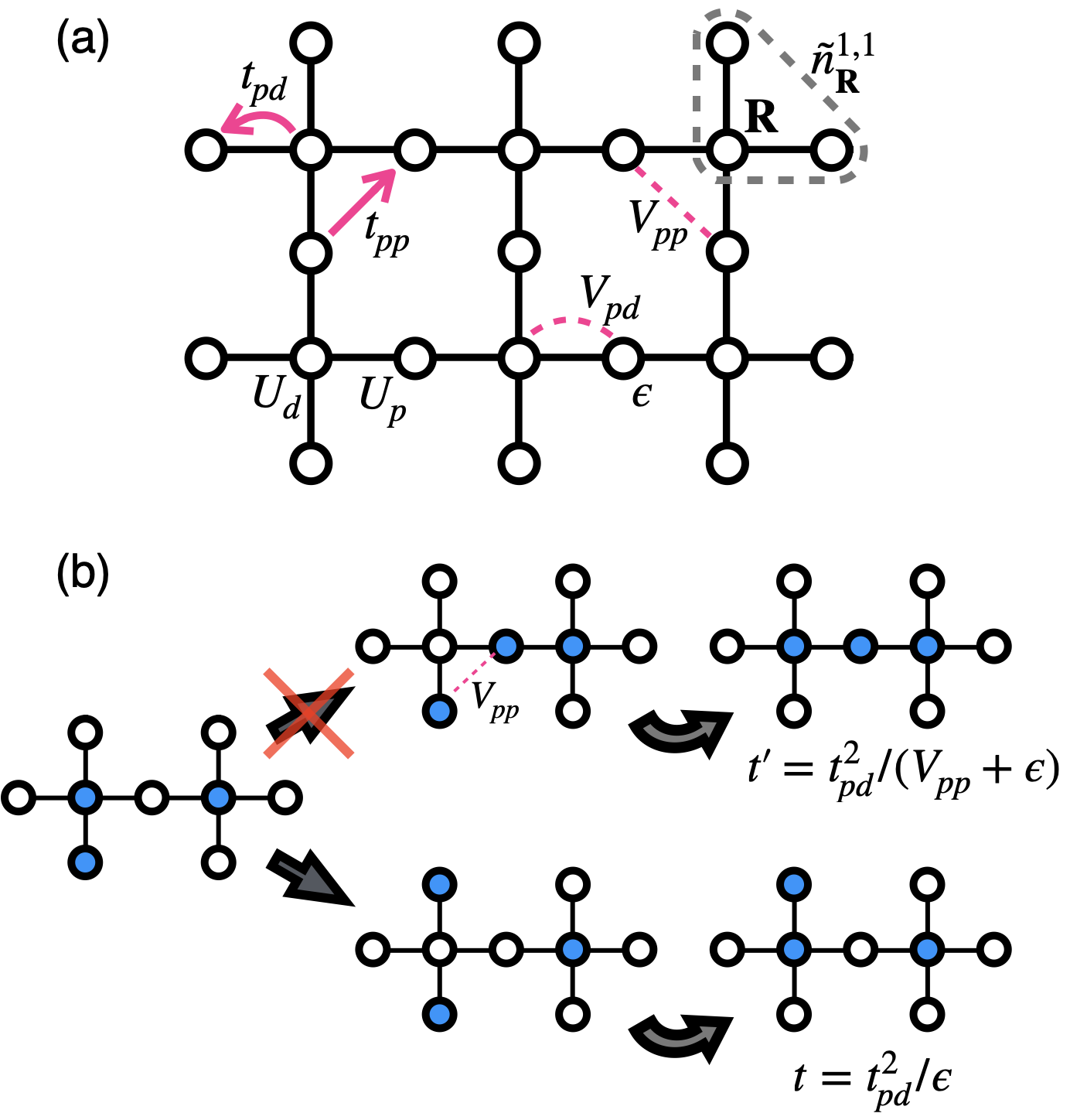}
    \caption{(a) Parameters in the Emery model as defined in the text around (\ref{eq:H_0}-\ref{eq:H_1}).
The hopping term $t_{pp}=0$ is neglected in this work.
    $\tilde n_{\v R}^{1,1}$ denotes the number of holes on a specified triangle (dashed gray line) as defined in \eqref{eq:triangle}. 
    (b) Two types of effective hopping processes in the semi-classical limit of the Emery model.
    The first process incurs a huge intermediate energy cost of $V_{pp}+\epsilon$, resulting in a negligible hopping term $t'$ in the strong-coupling limit.
    The second process is the only feasible process in the strong-coupling limit, granting holes quasi-1D mobility as described by \eqref{eq:effective Hamiltonian}. 
    }
    \label{fig:Emery model}
\end{figure}

Here, we review the derivation of an effective Hamiltonian closely related to \eqref{eq:solvable model}, obtained from the strong-coupling and  semi-classical limit of the Emery model, closely following Kivelson et al.  \cite{kivelson2004quasi1D}.
The Emery model was originally introduced to more accurately capture the local chemistry of cuprate superconductors than the simplified (single-band) Hubbard model 
\cite{emery1987highTc}. 
We will analyze this model independently of its initial motivation as a model for cuprate supercondutors. 
In the Emery model, defined on a Lieb lattice as in \eqref{eq:solvable model}, the elementary degrees of freedom are {\it holes}, with charge $e>0$ and spin $\frac 1 2$, rather than electrons, 
occupying Cu (O) sites on the vertices (bonds) of the Lieb lattice.

We define the hole creation operators on Cu and O sites as 
\begin{align}
    \tilde d^{\dagger}_{\v R,\sigma}\equiv d_{\v R,\sigma} \text{ and } \tilde p^{\dagger}_{\v R + \hat \delta /2,\sigma }\equiv p_{\v R+ \hat \delta /2,\sigma},
\end{align} 
respectively, where $\delta = x,y.$
Correspondingly, the hole number operators are defined as 
\begin{align}
    \tilde n_{\v R} &\equiv \sum_{\sigma} \tilde d^{\dagger}_{\v R,\sigma} \tilde d_{\v R,\sigma} = 2-n_{\v R},
    \\
    \tilde n_{\v R + \hat \delta /2 } &\equiv 
\sum_{\sigma}\tilde p_{\v R + \hat \delta /2,\sigma }^{\dagger} \tilde p_{\v R + \hat \delta /2,\sigma } = 2-  n_{\v R + \hat \delta /2 }.
\end{align}
Various terms in the model [illustrated in Fig. \ref{fig:Emery model} (a)] are defined as follows: 
\begin{itemize}
    \item[$(t)$] The hole hopping matrix element between the nearest-neighbor Cu and O sites is $t_{pd}$, while the hopping between two diagonally-neighboring O sites is $t_{pp}$ (which we set to zero in this work: $t_{pp} = 0$).
    \item[($\epsilon$)] The on-site potential energy for a hole occupying an O site is $\epsilon > 0$, measured relative to the potential energy on a Cu site.
    \item[($U$)] The on-site Coulomb repulsion between two {\it holes} occupying a Cu (O) site is $U_d$ ($U_p).$
    \item[$(V)$] The repulsion between two holes occupying nearest-neighbor Cu and O sites (two diagonally-neighboring O sites) is $V_{pd}$ ($V_{pp}$).
\end{itemize}
The Hamiltonian can then be written as a sum of two terms, $H_{\rm Emery}= H_0 + H_1$, where
\begin{align}
\label{eq:H_0}
    &H_0 =  \frac{U_d}{2} \sum_{\v R }\tilde n_{\v R} (\tilde n_{\v R} - 1 ) + \frac{U_p}{2} \sum_{\v R, \delta}\tilde n_{\v R + \frac {\hat \delta} 2} (\tilde n_{\v R+ \frac{\hat \delta} 2} - 1 ) 
    \nonumber \\
    &\!\!\!+
    V_{pd} \sum_{\v R}(\tilde n_{\v R} - 1) (\tilde n_{\v R + \frac {\hat x} 2} + \tilde n_{\v R + \frac {\hat y} 2} + \tilde n_{\v R - \frac {\hat x} 2} + \tilde n_{\v R - \frac {\hat y} 2} -2)
    \nonumber \\
    &\!\!\!+
    V_{pp}\sum_{\v R} \bigg ( \tilde n_{\v R + \frac{\hat x} 2} \tilde n_{\v R +\frac{\hat y} 2}  +
    \tilde n_{\v R - \frac{\hat x} 2} \tilde n_{\v R +  \frac{\hat y} 2}
    \nonumber \\
    & \ \ \ \ \ \ \ \ \ \ \ \ \ \ \ \ \ \ \ \ \ \ \ \ \ +
    \tilde n_{\v R -  \frac{\hat x} 2} \tilde n_{\v R -  \frac{\hat y} 2} +
     \tilde n_{\v R + \frac{\hat x} 2} \tilde n_{\v R -  \frac{\hat y}  2}
    \bigg )
\end{align}
and 
\begin{align}
\label{eq:H_1}
    &H_1 =
    \epsilon \sum_{\v R, \delta} \tilde n_{\v R + \frac {\hat \delta} 2} 
    \nonumber \\
    &
    +
    t_{pd }\sum_{\v R, \delta, \sigma} \bigg ( \tilde p^{\dagger}_{\v R + \frac{\hat \delta}{2}, \sigma} \tilde d_{\v R, \sigma}
    +
    \tilde d^{\dagger}_{\v R, \sigma} \tilde p_{\v R - \frac{\hat \delta} 2, \sigma}
    +
    {\rm H.c. }\bigg ).
\end{align}
Here, the sum $\delta$ is over bond flavors $\delta =x,y.$
The $V_{pd}$ term in $H_0$ includes an additional term, apart from the two-hole interaction term $\tilde n_{\v R} (\tilde n_{\v R + \frac {\hat x} 2} + \tilde n_{\v R + \frac {\hat y} 2} + \tilde n_{\v R - \frac {\hat x} 2} + \tilde n_{\v R - \frac {\hat y} 2})$, that is proportional to the total hole number $\tilde N$.
Since we will work in a fixed-number sector, this additional term is innocuous and is added for later convenience.
The $t_{pd}>0$ term has a positive sign as it represents {\it hole} hopping matrix element; however, this sign has no physical consequence as it can be flipped by a gauge transformation, $\tilde d_{\rm R, \sigma} \to -\tilde d_{\rm R, \sigma}.$

The `vacuum' of the model is defined as the state where each Cu and O site is doubly occupied by electrons, meaning that there are no holes.
The parent insulating state is defined as the state with one hole per square unit cell, which predominantly occupies Cu sites in the semi-classical limit $\epsilon \gg t$.
In the strong-coupling limit $U_d \gg \epsilon$ (with more precise conditions specified later), additional doped holes will occupy O sites.
We denote the  hole concentration per unit cell as 
\begin{align}
   1+  x= \frac 1 {L^2} \sum_{\v R}(\tilde n_{\v R} + \tilde n_{\v R+\hat x /2} + \tilde n_{\v R+\hat y /2}).
\end{align}
In the following, we consider the doping regime $0\leq x \leq 1$.

In the following, we review Ref. \cite{kivelson2004quasi1D}, which demonstrated that the Emery model supports the quasi-1D electron dynamics in the strong-coupling and semi-classical limit,
$U_d,U_p,V_{pd}, V_{pp} \gg \epsilon \gg t_{pd} \geq 0$. [Additional parameter constraints must also be satisfied, as specified below in \eqref{eq:strong coupling}].

First, it is convenient to define four triangles associated with each Cu site $\v R$ and two adjacent O sites [see Fig. \ref{fig:Emery model} (a)].
We label these triangles by $\{\v R, s,s' \}$, where $s,s' = \pm 1$ define the locations of two adjacent O sites, $\v R + \frac s 2\hat x $ and $\v R + \frac {s'} 2\hat y $.
By defining the hole number occupying such a triangle as
\begin{align}
\label{eq:triangle}
    \tilde n_{\v R}^{s,s'} \equiv \tilde n_{\v R} + \tilde n_{\v R +\frac s 2 \hat x}
    + \tilde n_{\v R +\frac {s'} 2 \hat y},
\end{align}
we can rewrite $H_0$ as \footnote{
The third term in \eqref{eq:H_0 projector} has a prefactor ${V_{pd}}/{4}$ that is different by a factor of $1/ 2$ from the expression (A5) of Ref. \cite{kivelson2004quasi1D}. We believe that this is a typo.
}
\begin{align}
\label{eq:H_0 projector}
     &H_0 =  \left ( \frac{U_d}{2} -V_{pd} \right )\sum_{\v R }\tilde n_{\v R} (\tilde n_{\v R} - 1 )
     + \left ( \frac{U_p}{2} -V_{pd} \right ) \times 
      \nonumber \\
     &
      \sum_{\v R, \delta}\tilde n_{\v R + \frac {\hat \delta} 2} (\tilde n_{\v R+ \frac{\hat \delta} 2} - 1 ) 
    +
    \frac {V_{pd}} 4  \sum_{\v R,s,s'}(\tilde n^{s,s'}_{\v R} - 1) (\tilde n^{s,s'}_{\v R} - 2)
    \nonumber \\
    &+
    \left ( V_{pp} - \frac{V_{pd}}{2} \right ) \sum_{\v R} \bigg ( \tilde n_{\v R + \frac{\hat x} 2} \tilde n_{\v R +\frac{\hat y} 2}  +
    \tilde n_{\v R - \frac{\hat x} 2} \tilde n_{\v R +  \frac{\hat y} 2}
    \nonumber \\
    & \ \ \ \ \ \ \ \ \ \ \ \ \ \ \ \ \ \ \ \ \ \ \ \ +
    \tilde n_{\v R -  \frac{\hat x} 2} \tilde n_{\v R -  \frac{\hat y} 2} +
     \tilde n_{\v R + \frac{\hat x} 2} \tilde n_{\v R -  \frac{\hat y}  2}
    \bigg ).
\end{align}
By restricting the parameter space to 
\begin{align}
    {U_d}/{2},\ {U_p}/{2},\ 2 V_{pp} > V_{pd}>0,
\end{align}
$H_0$ becomes positive semi-definite.
The zero-energy ground state manifold of $H_0$ consists of hole configurations satisfying the following conditions:
\begin{itemize}
    \item[(A)] No site is doubly occupied by holes.
    \item[(B)] Each triangle $\{\v R ,s,s' \}$ is  occupied by either one or two holes.
    \item[(C)] Two diagonally-neighboring O sites are not simultaneously occupied by holes. 
\end{itemize}
The strong-coupling limit is defined as the regime where the potential term $\epsilon>0$ is much smaller than the terms in \eqref{eq:H_0 projector}
\begin{align}
\label{eq:strong coupling}
    \frac {U_d}2 -V_{pd},\  \frac {U_p} 2 -V_{pd},\  V_{pd}, \ V_{pp}-\frac {V_{pd}} 2 \gg \epsilon
\end{align}
For $\epsilon > 0$, the zero-energy configurations are further restricted to those in which
\begin{itemize}
    \item[(D)] every Cu site is occupied by a single hole.
\end{itemize}

The above classical degeneracy is lifted by considering a quantum hopping term $t_{pd}>0$.
In the semi-classical limit, $U,V,\epsilon \gg t_{pd}$, there are two possible second-order processes by which a $p$-hole on a particular (say $y$) bond could move. 
[See Fig. \ref{fig:Emery model} (b) for an illustration.]
In these processes, the $d$-hole on the Cu site  neighboring the $p_y$-hole may move, in an intermediate step, to a neighboring O site on either $x$- or $y$-bond.
The $p_y$-hole could then move to fill the now-empty Cu site.
However, in the strong-coupling limit \eqref{eq:strong coupling}, the intermediate step where the $d$-hole occupies the $x$-bond is prohibited due to the hard-core constraint between the two diagonally-neighboring O sites (C).
The remaining second order process is captured by the following effective Hamiltonian
\begin{align}
\label{eq:effective Hamiltonian}
    H_{\rm eff} \!= \!
-t \sum_{\v R}
    \sum_{\delta = x,y}\sum_{\sigma,\sigma'  }
    &\left ( \tilde d^{\dagger}_{\v R,\sigma}  \tilde  p^{\dagger }_{\v R + \frac{\hat \delta } 2,\sigma'}  \tilde d_{\v R,\sigma'}   \tilde p_{\v R - \frac{\hat \delta } 2,\sigma}  
    \!
    +
    {\rm H.c.} \right ) \nonumber
     \\
     & + [U =\infty] + [V_{pp}=\infty],
\end{align}
where  $t =  t_{pd}^2/\epsilon$.
Here, it is implicitly assumed that each Cu site is always occupied by a single $d$-hole, whereas the hole concentration on $p$-sites per unit cell can vary $0\leq x \leq 1$.
The effective Hamiltonian $H_{\rm eff}$ is identical to \eqref{eq:solvable model} except for the last term, $[V_{pp}=\infty]$, which serves as a shorthand notation for projecting out hole configurations that violate the constraint (C).

\begin{figure}
    \centering
    \includegraphics[width=\linewidth]{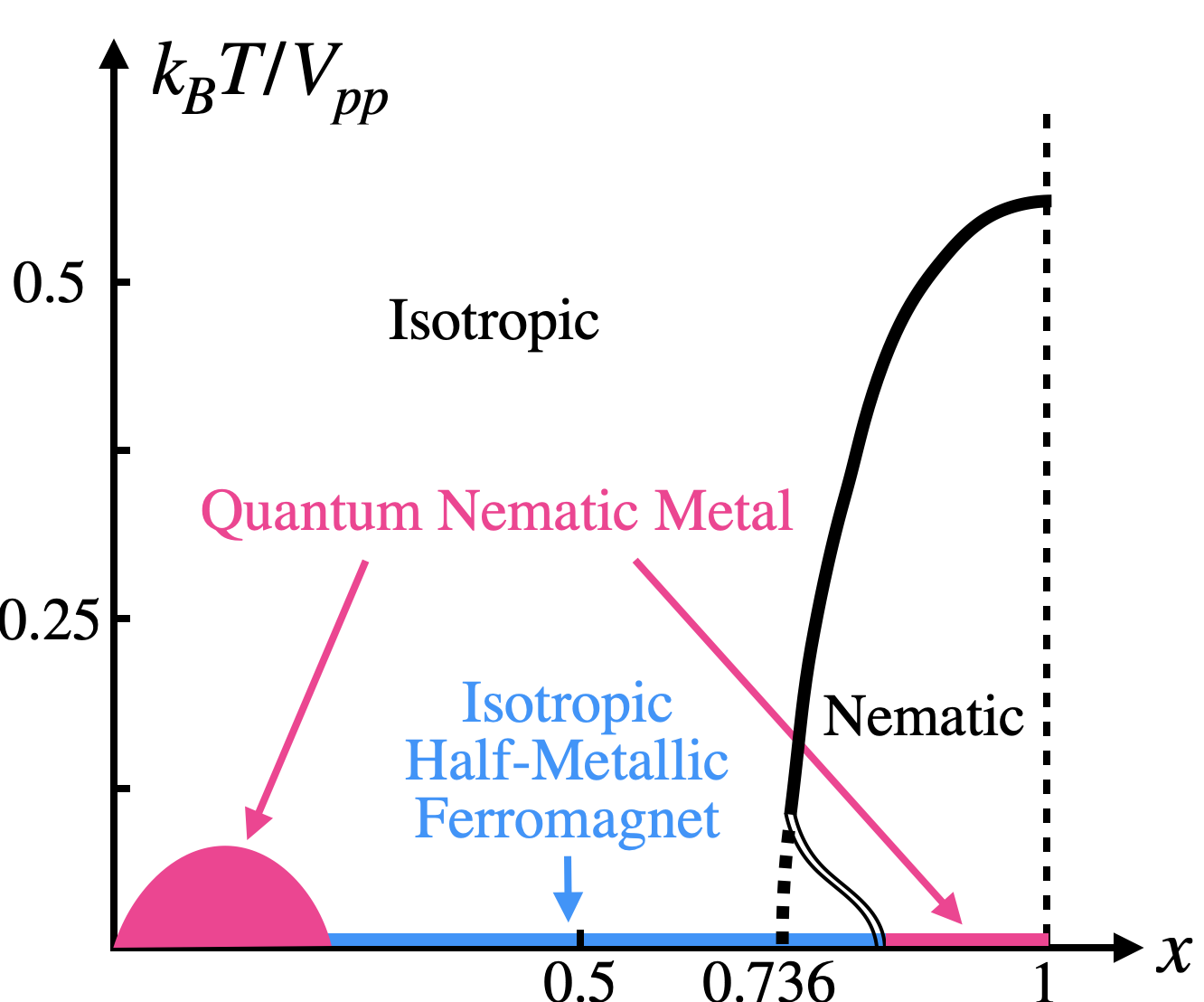}
    \caption{Classical (isotropic and nematic) and quantum (quantum nematic metal and isotropic half-metal) phases of the Emery model in the strong-coupling and semi-classical limit, described by $H_{\rm eff}$ \eqref{eq:effective Hamiltonian}.
    The black solid and dotted lines (taken from Ref. \cite{binder1980Ising}) mark the classical continuous nematic-isotropic phase boundary for $t=0$, which terminates at $x_c(T\to0) \approx 0.736$.
    Quantum effects [$t>0$] modify the finite-temperature phase diagram, as the isotropic phase may become energetically favorable over the nematic phase due to reduced Fermi pressure; 
    the double black line indicates this quantum-modified nematic-isotropic phase boundary, replacing the dotted line.
    Near $x=0$ and $x=1,$  quantum nematic phases are stabilized, becoming fully nematic at $T=0$ with complete spin degeneracy.
    Note that the distinction between quantum and classical nematic phases near $x=1$ is evident only at $T=0$, where any non-zero $t>0$ converts partial nematicity into full nematicity.
    In contrast, the nematicity near $x=0$ is purely a quantum effect.
    Away from these regions, the isotropic phase is expected to be the ground state and is characterized by full spin polarization at $T=0$.
    }
    \label{fig:Emery phase diagram}
\end{figure}

\smallskip

{\bf Classical phases of 
$H_{\rm eff}$  at $t=0$}.
Ignoring the spin degree of freedom and setting $t=0$, $H_{\rm eff}$ \eqref{eq:effective Hamiltonian} reduces to the hard-core lattice gas model with infinite nearest-neighbor repulsion on a square lattice (defined by the locations of $p$ orbitals) \cite{baxter2016exact}.
By relaxing the infinite repulsion condition [$V_{pp} =\infty$] to finite but large $V_{pp}$, the problem becomes equivalent to the square-lattice Ising antiferromagnet in a magnetic field, with $J=V_{pp}/4$ and magnetization per site $M =\frac 1 2 (1-x)$.

The phase diagram of this problem is well-known from the Monte Carlo  \cite{binder1980Ising} and transfer matrix calculations \cite{baxter1980hard, blote1990accurate}, which show that, in lattice gas terms, the high-temperature homogeneous liquid phase transitions to a
$(\pi,\pi)$ staggered crystal phase for $T<T_c(x)$ and $x\gtrsim0.736$ (see Fig. \ref{fig:Emery phase diagram}).
On the original Lieb lattice, this $(\pi,\pi)$ phase corresponds to the nematic phase with $\left < \tilde n_{\v R + \hat x /2 } \right > \neq \left < \tilde n_{\v R + \hat y /2 } \right >$.
In the zero temperature limit (or equivalently, $V_{pp}\to \infty$), the critical doping required for the existence of nematic phase is $x_c(T\to 0) \approx 0.736$ \cite{baxter1980hard}.
The phase transition is always continuous and falls within the Ising universality class \cite{binder1980Ising, baxter1980hard}. 
\footnote{
Ref. \cite{kivelson2004quasi1D} incorrectly claimed that, in addition to the isotropic and nematic phases, a region of two-phase coexistence exists at low temperature (see their Fig. 4).
Instead, the nematic-isotropic (or in Ising model terms, antiferromagnetic-paramagnetic) transition is always continuous at any finite temperature, and no two-phase region exists \cite{binder1980Ising,baxter1980hard,blote1990accurate}.
The transition occurs via the proliferation of nematic (or antiferromagnetic) domain walls \cite{muller1977interface}.
However, with an additional small next-nearest-neighbor (NNN) attraction (or NNN ferromagnetic coupling in the Ising terms), 
a two-phase region does indeed emerge at low temperatures \cite{binder1980Ising}. 
}
In the absence of quantumness ($t=0$), the nematic phase is characterized by partial nematicity, where both $x$- and $y$-bonds are occupied, albeit with some imbalance, except at $x=1$ and $T=0$.
At $x=1$ and $T=0$, the system becomes fully nematic, with only one bond flavor occupied.

\smallskip

{\bf Quantum phases 
of $H_{\rm eff}$}.
A small but finite $t>0$ modifies the low-temperature portion of the phase diagram discussed above.
Ref. \cite{kivelson2004quasi1D} argues that the {\it fully nematic} phase, characterized by occupation of only one bond flavor (say, $\left< n_{\v R + \hat x/2} \right > \neq  0$ and  $\left< n_{\v R + \hat y/2} \right > =0$), is  stable for $0< x\ll 1$ and $0< 1-x \ll 1$.
For completeness, we briefly repeat the argument for the dilute hole doping case $0<x\ll 1$.
In this case, the band structure \eqref{eq:band structure} can be analyzed within an effective-mass approximation: $E_k \approx -2 t +  t  k^2$.
The fully nematic phase then has the energy
\begin{align}
\label{eq:nematic}
    E_{\rm nem} = E_0 + \Delta_c x + \frac{\pi^2 t }{3}x^3 + O(x^5), 
\end{align}
where $\Delta_c = 2V_{pd} + \epsilon - 2t$. 
The cubic term is the contribution from Fermi pressure.
In contrast, an isotropic phase has reduced Fermi pressure but costs additional interaction energy due to $V_{pp}$ term:
\begin{align}
\label{eq:isotropic}
    E_{\rm iso} = E_0 + \Delta_c x + V_{\rm eff} x^2 + \frac 1 4 \frac{\pi^2 t }{3}x^3 + O(x^5). 
\end{align}
Here, $V_{\rm eff}$ represents a renormalized effective interaction between holes on intersecting rows and columns.
Therefore, we see that the nematic phase is stabilized for $0 < x < x_n \equiv \frac{4 V_{\rm eff}}{\pi^2 t} $ at $T=0$, while the isotropic phase becomes stable for $x_n<x$.

In the fully nematic phase, no nontrivial spin permutation can occur, as discussed previously, leading to fully degenerate spin states.
However, the isotropic ground state at intermediate doping exhibits full spin polarization due to the considerations in Sec. \ref{sec:Solvable model for a half-metal}.
\cite{footnote:Vpp}
Thus, {\it the quasi-1D electron dynamics exhibits competing tendencies toward orbital ferromagnetism (full nematicity) and spin ferromagnetism (full spin polarization)}. 
Full spin polarization arises whenever the system is {\it not} fully nematic;
however, in the fully nematic phase, all spin states are necessarily degenerate due to its full permutation nonergodicity.
Combining these considerations, we present a schematic phase diagram of the strong-coupling and semi-classical limit of the Emery model in Fig. \ref{fig:Emery phase diagram}.

\section{\textls[-20]{Quasi-1D vacancy in Wigner crystal}}
\label{sec:Wigner crystal}
In this section, we examine another example of quasi-1D particle originating from a different mechanism: a highly anisotropic lattice distortion (or phonon cloud) surrounding a quasi-particle. 
As an illustration, we consider a vacancy defect in a (triangular lattice) Wigner crystal (WC)---the ground state of a two-dimensional electron gas (2DEG)
\begin{align}
     \label{eq:2DEG}
    H = \sum_{i}\frac{\v p_i^{\, 2}}{2m} 
    + \sum_{i<j} \frac{e^2}{4\pi \epsilon} \frac{1}{|\v r_i -\v r_j|} ,
\end{align}
in the strong-coupling $r_s \to \infty$ limit. 
Here, $r_s =a_0/a_{\rm B}$ characterizes the ratio of Coulomb interaction strength to kinetic energy, where $a_0 = 1/\sqrt{\pi n}$ is the average interparticle distance, $n$ is the electron density, and $a_\mathrm{B} = 4\pi \epsilon \hbar^2 /m e^2$ is an effective Bohr radius.
We set the WC lattice constant to 1, with lattice sites indexed by $\v r = i\v e_1 + k \v e_3 $ ($i,k = 1,\cdots, L\equiv 0$), where $\v e_1 = [1,0]$ and $\v e_3 = [-1/2,\sqrt 3 /2]$.
We also define $\v e_2 = [1/2,\sqrt 3 /2]$ for later convenience.

\begin{figure}
    \centering
    \includegraphics[width=\linewidth]{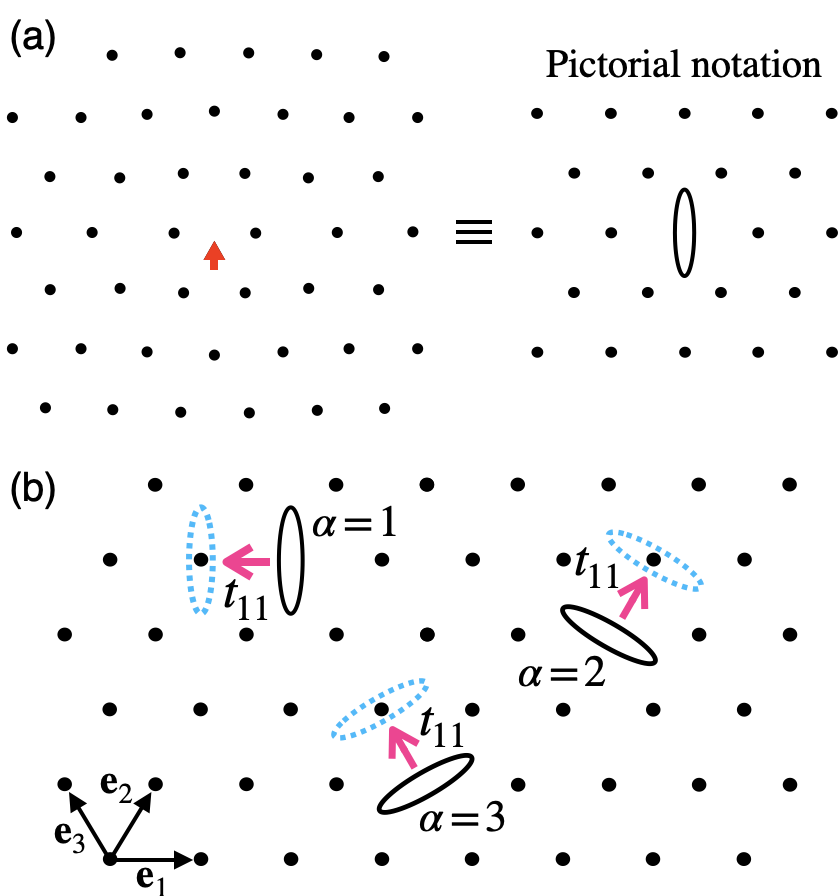}
    \caption{(a) Classical configuration of a vacancy (centered at the position indicated by the red arrow).
    The classical vacancy configuration, obtained by minimizing the Coulomb interaction in the presence of a single vacancy, breaks $C_3$ symmetry, resulting in three distinct orientations (or orbitals) $\alpha =1,2,3$ for vacancies. 
    (b) The vacancy dynamics is predominantly one-dimensional in the semi-classical $r_s \to \infty$ limit and is governed by the $t_{11}$ process \eqref{eq:vacancy}.
    Here, upon applying the $t_{11}$ term, vacancies initially at the black ovals are moved to the cyan ovals.
    Figures are adapted from Ref. \cite{kim2023dynamical}.
    }
\label{fig:vacancy_configuration}
\end{figure}

The classical vacancy defect of the WC is known to have reduced symmetry $D_2$, rather than the $D_6$ of the underlying WC, resulting in three inequivalent vacancy orientations (or orbitals), $\alpha=1,2,3$, which are related by $C_3$ rotations [see Fig. \ref{fig:vacancy_configuration} (a)] \cite{cockayne1991energetics, kim2023dynamical}.
The vacancy creation operator is defined as follows.
Let $f_{\v r,\sigma}$ be the fermionic operator that annihilates an electron at site $\v r$ and  $b_{\v r,\alpha}^{\dagger}$ be the (hard-core) bosonic operator that relaxes the nearby WC electrons into the vacancy configuration specified by $\alpha$, which  minimizes the classical Coulomb energy [Fig. \ref{fig:vacancy_configuration} (a)].
A vacancy creation operator can then be expressed as $f_{\v r,\sigma}b^{\dagger}_{\v r,\alpha}$, which first removes an electron (via $f_{\v r,\sigma}$) and dresses it with a polaronic cloud (via $b^{\dagger }_{\v r,\alpha}$).

The quantum hopping and exchange processes involving such an anisotropic vacancy were studied by the first author in Ref. \cite{kim2023dynamical} within the semi-classical instanton approximation [See their Fig. 4 (d-e) and Eqs. 16 \& 20].
It was shown that in the asymptotic $r_s \to \infty$ limit (specifically, when $r_s \gtrsim 320$), the vacancy dynamics becomes predominantly one-dimensional, described by $t_{11}$ process  illustrated in Fig. \ref{fig:vacancy_configuration} (b), as the tunnel barrier for this process becomes lower than for any other processes. 
\footnote{
For further discussion on the regime of such 1D mobility, see Footnote [27] in Ref. \cite{kim2023dynamical}.
Achieving such a regime of $r_s$ experimentally may be  challenging; exploring  more experimentally relevant platforms that host such `lineonic' defect originating from polaronic effects is an interesting future direction.
}
The effective Hamiltonian governing such a 1D vacancy dynamics is 
\begin{align}
\label{eq:vacancy}
      H^{\rm v}_{\rm eff} = -t_{11} \sum_{\v r}
    \sum_{\alpha= 1}^3\sum_{\sigma} & \left (  f^{\dagger}_{\v r + \v  e_\alpha,\sigma}  f_{\v r,\sigma} b^{\dagger}_{\v r,\alpha}b_{\v r + \v e_\alpha,\alpha} + {\rm H.c.} \right ) \nonumber \\
    & \ \ \ \ \  \ \ \ \ \ \ \ \ \ \  + [U =\infty],
\end{align}
describing a vacancy hopping from site  $\v r+\v e_{\alpha}$ to $\v r $.
The number operator for $\alpha$-vacancies at site $\v r$ will be denoted by $n^{\rm v}_{\alpha}(\v r) =  b^{\dagger }_{\v r, \alpha} b_{\v r, \alpha}.$
Note that, due to the hard-core constraint, $\sum_{\alpha}n^{\rm v}_{\v r, \alpha} +\sum_{\sigma } f^{\dagger }_{\v r, \sigma} f_{\v r, \sigma} =1$ is satisfied for each $\v r$.
We study this problem \eqref{eq:vacancy} in an  $L \times L $ effective triangular lattice.

\smallskip
\noindent
{\bf Subsystem Symmetry on $i$th $\alpha$-line.}
Due to the 1D mobility of vacancies, the model exhibits subsystem symmetries analogous to \eqref{eq:conservation_Emery1} and \eqref{eq:conservation_Emery2} \cite{footnote:subsystem}.
Specifically, the number of $\alpha$-vacancies on each line parallel to $ \v e _{\alpha}$  is conserved:
\begin{align}
\label{eq:subsystem symmetry 1 for WC}
N^{\rm  v}_{1 }(k) &\equiv \sum_{i=1}^L n^{\rm v}_{1}({i \v e_1 + k \v e_3 }),\ \ {\rm for }\ \ k = 1,\cdots,L,
\\
\label{eq:subsystem symmetry 2 for WC}
 N^{\rm  v}_{2}(i) &\equiv \sum_{j=1}^L n^{\rm v}_{2 }({i \v e_1 + j \v e_2 }),
 \ \ {\rm for }\ \ i = 1,\cdots,L,
 \\
\label{eq:subsystem symmetry 3 for WC}
N^{\rm  v}_{3}(i) &\equiv \sum_{k=1}^L n^{\rm v}_{3}({i \v e_1 + k \v e_3 }), \ \ {\rm for }\ \ i = 1,\cdots,L.
\end{align}
The lines along which the number of $\alpha$-th vacancies is counted in (\ref{eq:subsystem symmetry 1 for WC}-\ref{eq:subsystem symmetry 3 for WC}) will be referred to as the {\it $i$th $\alpha$-line.}
The total number of $\alpha$-vacancies will be denoted by
\begin{align}
    N^{\rm  v}_{\alpha,{\rm tot}} \equiv \sum_{i=1}^L N^{\rm v}_{\alpha}(i).
\end{align}

\smallskip

\begin{figure}[t]
    \centering
    \includegraphics[width=\linewidth]{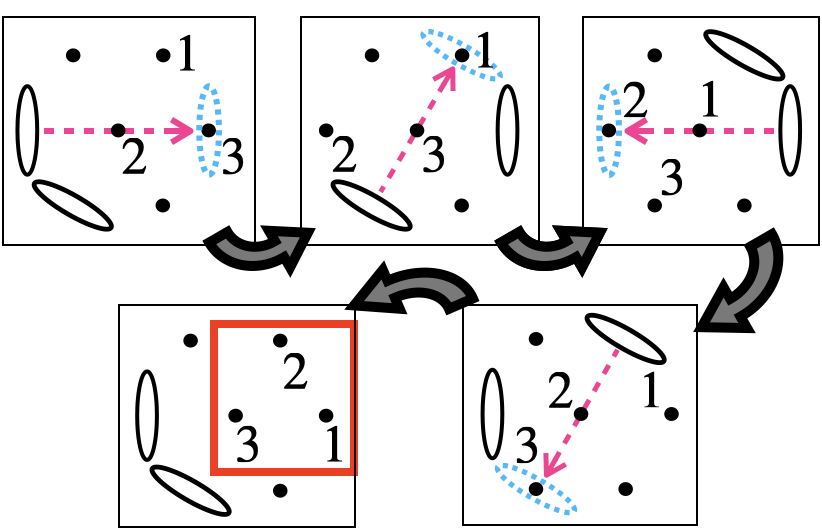}
    \caption{Elementary 3-cycle (red box) induced by the correlated motions of two vacancies.
    Compare the three numbered electrons in the first panel (upper left) to those in the final panel (lower left).
    In each step, a vacancy moves along the dashed magenta arrow to the site indicated by the dashed cyan oval.
    }
    \label{fig:vacancy exchange}
\end{figure}

The correlated motions of 1D vacancies can induce magnetic correlations through a nontrivial multi-spin ring exchange among WC electrons.
We illustrate the simplest such process in Fig. \ref{fig:vacancy exchange}, where an {\it elementary 3-cycle} (red box) is induced by the correlated motions of two vacancies.
This process is analogous to the elementary 3-cycle in the solvable model \eqref{eq:solvable model} shown in Fig. \ref{fig:solvable Lieb lattice} (b).
The following two results (analogous to Propositions \ref{prop:evenness for Lieb} and Theorem \ref{thm:exact ferro ground state}) demonstrate that any nontrivial multi-spin ring exchange induced by 1D vacancies is of even parity, thereby mediating ferromagnetism in accordance with the Thouless rule.

\smallskip

\begin{proposition}
\label{prop:evenness for WC}
\rm{\bf $\pmb{[}$Evenness of Ring Exchange$\pmb{]}$}
Consider the Hamiltonian \eqref{eq:vacancy} on an $L\times L $ system with either the open (OBC) or periodic (PBC). 
For the PBC, we additionally require the following condition:
\begin{itemize}
    \item[(a)] The number of $\alpha$-vacancies on each $\alpha$-line is odd (even) for even (odd) $L$, unless there are no $\alpha$-vacancies on that line. In other words, $L- N^{\rm v}_{\alpha}(i)$ is odd, so that $(-1)^{L-N^{\rm v}_{\alpha}(i)-1} =1 $ for each $i$ and $\alpha$, unless $N^{\rm v}_{\alpha}(i) = 0$.
\end{itemize}
Then, any permutation $\pi$ of $N$ Boltzmannian electrons induced by the Hamiltonian dynamics (\ref{eq:vacancy}) is of even parity: ${\rm sgn}(\pi) = 1$.
\end{proposition}

\begin{theorem}
\label{thm:exact ferro GS for vacancy}
\rm{\bf $\pmb{[}$Itinerant Ferromagnetism from 1D Vacancies$\pmb{]}$}
The Hamiltonian \eqref{eq:vacancy} on an $L\times L$ system supports a fully spin-polarized state as one of its ground states, provided that the conditions of Proposition \ref{prop:evenness for WC} are met.
\end{theorem}

{\it Proof of Proposition \ref{prop:evenness for WC}.}
The proof is a slight variation of the one given for Proposition \ref{prop:evenness for Lieb}.
Let $({\pmb{\mathcal R}}_i) \equiv ({\pmb{\mathcal R}}_1, \cdots, {\pmb{\mathcal R}}_{N_{\rm site}})$ represent a particular electron-vacancy configuration, where electron and vacancy indices are given by  $i=1,\cdots,N$ and $i=N+1,\cdots,N_{\rm site}$, respectively.
The configuration of electron spins and vacancy orientations are specified by $\sigma_i$ and $\alpha_i$, respectively.
As before, any other electron-vacancy configuration $(\pmb{\mathcal{R}}_i')$ defines a permutation $\Pi\in {\rm S}_{N_{\rm site}}$ via $\pmb{\mathcal{R}}_i' = \pmb{\mathcal{R}}_{\Pi(i)}$.
Each term of $H_{\rm eff}$, when applied to a particular electron-vacancy configuration, either annihilates that configuration or yields a new configuration.
In the latter case, such a term acts as an {\it elementary transposition} involving one electron and one vacancy.

Due to the strict 1D mobility of vacancies, any configuration $({\pmb{\mathcal R}}_i')$ connected to $({\pmb{\mathcal R}}_i)$ through repeated applications of $H^{\rm v}_{\rm eff}$ has each vacancy ${\pmb{\mathcal R}}_i'$ $(i=N+1,\cdots,N_{\rm site})$ positioned at one-dimensional distances $\{d_i|\ i=N+1,\cdots,N_{\rm site}\}$ along the corresponding $\alpha_i$-line relative to their initial positions $({\pmb{\mathcal R}}_i)$.
Here, the distance $d_i$ is measured in the positive $\v e_{\alpha_i}$ direction.

It is possible to show that the following quantity is invariant under the dynamics of \eqref{eq:vacancy} 
\footnote{
This invariant is reminiscent of the invariant used in proving the unsolvability of the 15-puzzle.
}:
\begin{align}
\label{eq:invariant for vacancy}
    I({\pmb{\mathcal R}}_1',\cdots,{\pmb{\mathcal R}}'_{N_{\rm site}}) \equiv {\rm sgn}(\Pi)\cdot (-1)^{\sum_{i=N+1}^{N_{\rm site}} d_i},
\end{align}
where $\Pi$ is the permutation defined relative to $({\pmb{\mathcal R}}_i)$.
It suffices to observe that every elementary transposition leaves $I$ invariant.
Each elementary transposition simultaneously changes the parity ${\rm sgn}(\Pi)$ by $-1$ and one of the distance factor $(-1)^{d_i}$ by $-1$, leaving $I$ unchanged.
Since $I({\pmb{\mathcal R}}_i)=1$ for the initial configuration, we have $I({\pmb{\mathcal R}}_i')=1$ for any configuration $({\pmb{\mathcal R}}_i')$ connected to $({\pmb{\mathcal R}}_i)$ by the dynamics of \eqref{eq:vacancy}.

Now, consider a final electron-vacancy configuration $({\pmb{\mathcal R}}_i')$ connected to $({\pmb{\mathcal R}}_i)$ by the dynamics of \eqref{eq:vacancy}, where electron positions permuted only among themselves. 
The associated permutation $\Pi$  in this case can be decomposed as the product of permutation involving only electrons ($\pi$) and only vacancies ($\pi_{\rm v}$): $\Pi = \pi \cdot \pi_{\rm v}.$
In the meantime, since each (Boltzmannian) vacancy is strictly confined to the line associated with its orientation $\alpha_i$, its final position  for the OBC must be the same as in the initial configuration; that is, ${\pmb{\mathcal R}}_i'  =  {\pmb{\mathcal R}}_i$ and $d_i = 0$ for $i=N+1,\cdots, N_{\rm site}$.
This implies ${\rm sgn}(\pi_{\rm v})=1$, and thus $ 1= I({\pmb{\mathcal R}}_i') = {\rm sgn}(\pi)$, completing the proof for the OBC.
For the PBC, it is possible for $\alpha $-vacancies on each $\alpha$-line to undergo cyclic permutations.
In such cases, $I$ acquires a factor of $(-1)^L$ [from the distance factor in \eqref{eq:invariant for vacancy}] in addition to a factor of $(-1)^{N^{\rm v}_{\alpha}(i)+1}$ [from the sign factor], where $N^{\rm v}_{\alpha}(i)$ is the number of $\alpha $-vacancies on the $\alpha$-line in which the cyclic permutation occurs.
By condition (a), the combined factor $(-1)^{L-N^{\rm v}_{\alpha}(i)-1} =1.$
Therefore, $1= I({\pmb{\mathcal R}}_i') = {\rm sgn }(\pi) $, completing the proof for both boundary conditions.
$\square$

\smallskip

{\it Proof of Theorem \ref{thm:exact ferro GS for vacancy}.}
As in the proof of Theorem \ref{thm:exact ferro ground state}, we aim to identify a Perron-Frobenius basis in which all off-diagonal matrix elements are non-positive.
This can be accomplished if the bosonic operator $b$ could be transmuted into a fermionic operator. 
Such transformation would eliminate the fermion sign problem, making the system effectively `bosonic.'

\begin{figure}
    \centering
    \includegraphics[width=\linewidth]{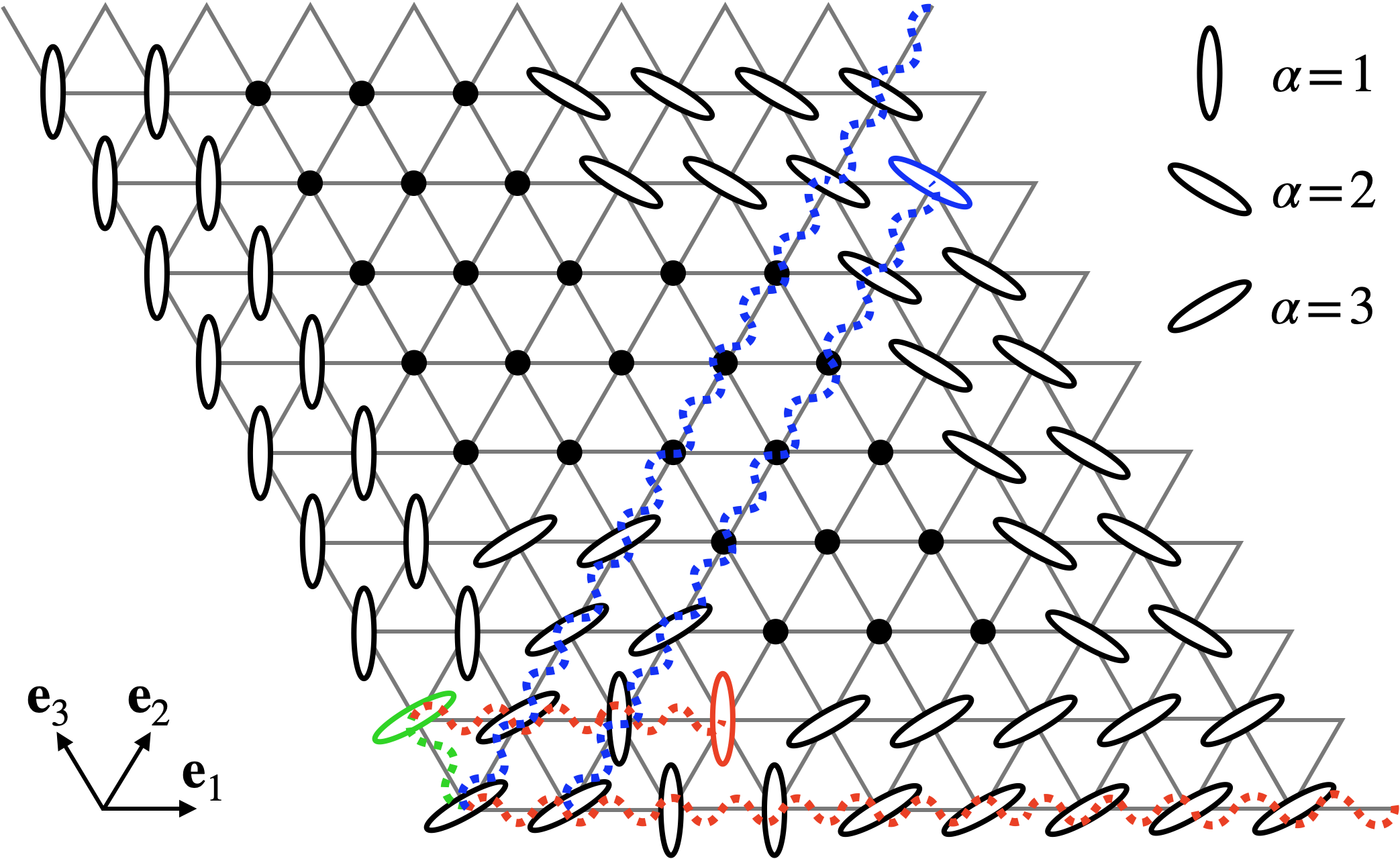}
    \caption{Canonical vacancy configuration and Jordan-Wigner fermionization of vacancies on a $9 \times 9$ system. 
    A canonical vacancy configuration, defined for a periodic $L \times L$ system that satisfies conditions (a-d) of Proposition \ref{prop:ergodicity for WC}, places vacancies of each orientation at sites specified by (\ref{eq:alpha1-1}-\ref{eq:alpha3-3}).
    Here, the site at the bottom-left corner is given by $\v e_1 + \v e_3.$
    Sites not occupied by a vacancy are occupied by an electron (black dot).
    In this arrangement,  $\alpha =1,3$ vacancies are placed as compactly as possible to the left and bottom and $\alpha =2$ vacancies are placed in the upper right corner.
    The topmost and rightmost sites must be identified with the bottommost and leftmost sites, respectively, to maintain periodicity.
    The even-permutation ergodicity (Proposition \ref{prop:ergodicity for WC}) and the resulting unique ferromagnetic ground state (Theorem \ref{thm:uniqueness for WC}) are demonstrated in each sector connected to this configuration. 
    The Jordan-Wigner strings (\ref{eq:JWstring1}-\ref{eq:JWstring3}), represented by the dotted curvy lines (red, blue and green), are attached to the corresponding types of vacancies (red, blue and green) to fermionize them.
    }
    \label{fig:JW transformation}
\end{figure}

Given the 1D mobility of vacancies, this  can be achieved through a Jordan-Wigner (JW) transformation, where JW strings are attached in a specific way for each $\alpha$-vacancy. 
For each site $\v r \equiv i \v e_1 + j \v e_2 = i \v e_1 + k \v e_3$, we define
\begin{align}
\label{eq:JW transformation}
    & c_{\v r, \alpha} \equiv (-1)^{N -  \sum_{\alpha' <\alpha} N^{\rm v}_{\alpha', {\rm tot}} } F_{\alpha}(\v r) b_{\v r, \alpha},
\end{align}
where $N$ is the total number of electrons, and 
\begin{align}
    \label{eq:JWstring1}
    &\!\!\!\!\!F_1(i \v e_1 + k \v e_3) \!  \equiv \! (-1)^ {i - \sum_{k'<k} \! 
    N^{\rm v}_{1}(k')
    -
    \sum_{i'<i}
    n^{\rm v}_1(i' \v e_1 + k \v e_3) }
    \\
     \!\!\!\!\!
         \label{eq:JWstring2}
     &\!\!\!\!\!F_2(i \v e_1 + j \v e_2) \!\equiv\! (-1)^ {j- \sum_{i'<i} 
    N^{\rm v}_{2}(i')
    -
    \sum_{j'<j}
    n^{\rm v}_1(i \v e_1 + j' \v e_2) }
    \\
        \label{eq:JWstring3}
     &\!\!\!\!\! F_3(i \v e_1 + k \v e_3) \!\equiv\! (-1)^ {k - \sum_{i'<i}\! 
    N^{\rm v}_{3}(i')
    -
    \sum_{k'<k}\!
    n^{\rm v}_1(i \v e_1 + k' \v e_3) }
\end{align}
See Fig. \ref{fig:JW transformation} for an illustration of these JW strings.
The $(-1)^{N}$ factor in \eqref{eq:JW transformation} ensures the correct fermionic anti-commutation relation between $c$ and $f$ operators, while the $(-1)^ {-\sum_{\alpha' <\alpha} N^{\rm v}_{\alpha', {\rm tot}} }$ factor ensures the anti-commutation relation between $c$ operators of different orientations $\alpha$.
Each $F_{\alpha}$ includes an additional factor $(-1)^{i}$, $(-1)^{j}$ or $(-1)^{k}$ besides the usual JW string;
as we will see, this factor flips the sign of the hopping matrix element, $t_{11}\to -t_{11}$, ensuring that the Hamiltonian matrix is non-positive. 
It is straightforward to verify $f$ and $c$ operators follow the usual fermionic anti-commutation relations.
The Hamiltonian in terms of the fermionic vacancy operators $c$ becomes \footnote{For the PBC case, the transformed Hamiltonian \eqref{eq:fermionic vacancy} maintains the usual periodic boundary condition
\begin{align*}
   \ \ \ \ c_{i \v e_1 +k\v e_3 + L \v e_{1}, 1} =
    (-1)^{L - N^{\rm v}_{1}(k)-1 } c_{i \v e_1 +k\v e_3, 1}=c_{i \v e_1 +k\v e_3, 1},
    \end{align*}
    and similarly 
\begin{align*}
    &c_{i \v e_1 +j\v e_2 + L \v e_{2}, 2} 
    =c_{i \v e_1 +j\v e_2, 2},
    \\
    &c_{i \v e_1 +k\v e_3 + L \v e_{3}, 3} 
    =c_{i \v e_1 +k\v e_3, 3}. 
\end{align*}
The condition that each $L- N^{\rm v}_{\alpha}(i)$ is odd is essential for preserving the PBC in \eqref{eq:fermionic vacancy}.
If it were even, the Hamiltonian \eqref{eq:fermionic vacancy} would instead exhibit an antiperiodic boundary condition, with a $\pi$-flux introduced along the corresponding $\alpha$-line.}
\begin{align}
\label{eq:fermionic vacancy}
      H^{\rm v}_{\rm eff} = + t_{11} \sum_{\v r}
    \sum_{\alpha= 1}^3\sum_{\sigma} & \left (  f^{\dagger}_{\v r + \v  e_\alpha,\sigma}  f_{\v r,\sigma} c^{\dagger}_{\v r,\alpha} c_{\v r + \v e_\alpha,\alpha} + {\rm H.c.} \right ) \nonumber \\
    &\ \ \ \ \  \ \ \ \ \ \ \ \ \ \ + [U =\infty].
\end{align}
In this form of the Hamiltonian \eqref{eq:fermionic vacancy}, each site is occupied by exactly one fermion, either of type $f$ or $c$, and 
$H^{\rm v}_{\rm eff}$ can be interpreted as a two-particle exchange operator between two sites.

Let us now introduce the Perron-Frobenius basis as follows.
First, we arbitrarily order the site from $\v r_1$ to $\v r_{L^2}$, with each site occupied by one $f$- or $c$-fermion
\begin{align}
\label{eq:basis for vacancies}
    \left |\{\sigma/\alpha \} \right > 
    \equiv 
    \cdots c^{\dagger}_{\v r_{i}, \alpha_i} \cdots f^{\dagger}_{\v r_{j}, \sigma_j} \cdots
    \vacuum,
\end{align}
where operators are arranged in increasing order of site indices from $1$ to $L^2$, with $i<j$.
Then, any non-zero off-diagonal matrix element is negative:
\begin{align}
\label{eq:matrix element for vacancy}
     \left <\{ \sigma/\alpha \} \right | H^{\rm v}_{\rm eff} \left |\{ \sigma'/\alpha' \} \right > = -t_{11}. 
\end{align}
This establishes that the Hamiltonian matrix in the basis \eqref{eq:basis for vacancies} is non-positive.

The fact that the fully spin-polarized state is one of the ground states can be shown by constructing a trial wave function analogous to \eqref{eq:trial ferromagnetic state for $T$} that is used in the proof of Theorem \ref{thm:exact ferro ground state}. $\square$

\smallskip

In proving the even-permutation ergodicity in Propositions \ref{prop:permutation ergodicity OBC} and \ref{prop:permutation ergodicity PBC},  and the consequent uniqueness of a ferromagnetic ground state in Theorem \ref{thm:uniqueness for Lieb}, we relied on the fact that the positional ergodicity holds for any electronic configuration under the dynamics of \eqref{eq:solvable model}.
However, positional ergodicity no longer holds for the Hamiltonian \eqref{eq:vacancy} due to the on-site hard-core interactions between vacancies, which are absent in \eqref{eq:solvable model}. 
For example, if two rows (parallel to $\v e_1$) are completely filled by $\alpha =1 $ vacancies, any vacancies or electrons interior to those lines are unable to move out, trivially breaking positional ergodicity.
However, determining whether positional ergodicity is satisfied within a subsector of the Hilbert space  defined by $N^{\rm v}_{\alpha}(i)$ (\ref{eq:subsystem symmetry 1 for WC}-\ref{eq:subsystem symmetry 3 for WC}) for more complicated vacancy configurations is much more challenging.

Therefore, we  restrict our analysis to sectors of the Hilbert space where the number of $\alpha$-vacancies is identical on each $\alpha$-line: i.e., $N^{\rm v}_{\alpha}(i)\equiv m_{\alpha}$ for $i=1,\cdots,L$ and $\alpha = 1,2,3$.
The total number of vacancies is then 
\begin{align}
\label{eq:total vacancy number}
    N^{\rm v}_{\rm tot} = \sum_{\alpha=1}^3 N^{\rm v}_{\alpha, {\rm tot}} =(m_1+m_2+m_3)L.
\end{align}
The following proposition establishes the even-permutation ergodicity within these sectors when vacancies of two or more distinct orientations are present.
This, in turn, implies the uniqueness of a ferromagnetic ground state as stated in Theorem \ref{thm:uniqueness for WC}.
The proof of this proposition is somewhat technical and is provided in Appendix \ref{app:ergodicity for WC}.

\begin{proposition}
\label{prop:ergodicity for WC}
{\rm{\bf{$\pmb{[}$Even-Permutation Ergodicity$\pmb{]}$}}}
Consider the Hamiltonian \eqref{eq:vacancy} on an $L \times L$ triangular lattice with the periodic boundary condition (PBC). 
We require that 
\begin{itemize}
    \item[(a)] the number of $\alpha$-vacancies on each $\alpha$-line is the same: $ N^{\rm v}_{\alpha}(i) \equiv m_{\alpha}$ for each $i$;
    \item[(b)] $L \!-   m_{\alpha}$ is odd unless $m_{\alpha }\! =\!0$ [as in Proposition \ref{prop:evenness for WC}];
    \item[(c)] there exist at least two 
    vacancy orientations; 
    if only two orientations are present, we also require that $2m_{\alpha} \leq L-  2$ for each $\alpha$;
    and
    \item[(d)] $\sum_{\alpha=1}^3 m_{\alpha} \equiv m \leq L-2$.
\end{itemize}
Then, the even-permutation ergodicity is satisfied under the 1D vacancy dynamics \eqref{eq:vacancy} within the sector connected to the {canonical vacancy configuration} as defined below.
\end{proposition}

A {\it canonical vacancy configuration} is defined as follows for a periodic $L\times L $ system that satisfies conditions (a-d) above.
See Fig. \ref{fig:JW transformation} for an illustration.
Without loss of generality, let us assume $m_3 \geq m_1 \geq 1 $ if only two vacancy orientations are present,
and  $m_2 \geq m_3 \geq m_1\geq 1 $ if all three orientations are present.
We first place $\alpha =1$ vacancies at sites 
\begin{align}
\label{eq:alpha1-1}
    \v e_1 + k \v e_3, 2\v e_1 + k \v e_3, \cdots, 
     m_1\v e_1 + k \v e_3
    \end{align}
for $m_3 + 1\leq k \leq L$ and at 
\begin{align}
\label{eq:alpha1-2}
    (m_1\!+\!1)\v e_1 \!+\! k \v e_3,(m_1\!+\!2)\v e_1 \!+\! k \v e_3 ,\cdots
    , 2m_1 \v e_1 \!+\! k \v e_3
\end{align}
for $ 1\leq k \leq m_3 $.
Then, $\alpha = 3$ vacancies are placed at
\begin{align}
\label{eq:alpha2-1}
    i \v  e_1 +  \v e_3, i\v e_1 + 2 \v e_3, \cdots, 
     i\v e_1 + m_3 \v e_3
    \end{align}
for $1\leq i \leq m_1$ and $2m_1 + 1\leq i \leq L$ and at 
\begin{align}
\label{eq:alpha2-2}
    i \v  e_1 \!+  (m_3\!+\!1) \v e_3, i\v e_1 \!+ (m_3\!+\!2) \v e_3, \cdots, 
     i\v e_1 \!+ 2m_3 \v e_3
\end{align}
for $m_1 + 1 \leq i \leq 2 m_1 $.
Finally, $\alpha = 2$ vacancies are placed at 
\begin{align}
\label{eq:alpha3-1}
    &(i-1) \v  e_1 +  (L-m_2-i+2) \v e_2, \cdots, 
    \nonumber \\
     &\ \ \ \ \ \ \ \ \ \ \ \ \ \ \ \ \ \ \ \ \ \ \ \ \ \ \ \ \ \ (i-1)\v e_1 + (L-i+1) \v e_2
\end{align}
for $1 \leq i\leq L-m_2 -m_3 + 1$,
at
\begin{align}
\label{eq:alpha3-2}
    & (i\!-\!1)\v  e_1 \!+\!  (m_3+1) \v e_2, \cdots, 
     (i\!-\!1)\v e_1\! +\! (L+1-i) \v e_2,
     \\
    & (i\!-\!1) \v  e_1 \!+ \! (2L\!- m_2 -\!m_3-i+2) \v e_2, \cdots, 
     (i\!-\!1)\v e_1 \!+\! L \v e_2
\end{align}
for $L-m_2-m_3+2 \leq i\leq L-m_3$,
and at
\begin{align}
\label{eq:alpha3-3}
    (i-1) \v  e_1 +  (L-m_2+1) \v e_2, \cdots, 
     (i-1)\v e_1 + L \v e_2
\end{align}
for $L-m_3+1 \leq i\leq L$.
Through this placement, the $\alpha = 1,3$ vacancies are positioned as compactly as possible toward the left and bottom, while $\alpha =2$ vacancies occupy the upper right corner, as shown in Fig. \ref{fig:JW transformation}.
\smallskip

Condition (a) of Proposition \ref{prop:ergodicity for WC} is assumed for convenience; although rigorous results beyond condition (a) were not obtained, we believe that the even-permutation ergodicity holds in a much wider range of sectors.
Condition (b) ensures the evenness of permutation, as established in Proposition \ref{prop:evenness for WC}, while conditions (c) and (d) ensure that any such even permutation can be obtained.
An additional condition for (c) in the presence of only two vacancy orientations ensures that vacancy configurations can be made canonical; 
if the condition $2m_{\alpha}\leq L-2$ is not met, a more complicated canonical vacancy configuration must be defined to demonstrate the even-permutation ergodicity.
The range of vacancy doping per site covered by the proposition is  
\begin{align}
\label{eq:vacancy doping range}
    2/L \leq \nu^{\rm v }\equiv N^{\rm v}_{\rm tot}/L^2 \leq (L-2)/L,
\end{align} 
which spans the full range of doping in the thermodynamic limit.
\footnote{
As in the strong-coupling limit of the Emery model \eqref{eq:effective Hamiltonian}, determining the global ground state for a given filling $\nu^{\rm v}$ requires comparing the ground state energies across different sectors of (\ref{eq:subsystem symmetry 1 for WC}-\ref{eq:subsystem symmetry 3 for WC}).
As discussed in Sec. \ref{sec:Emery model}, strong on-site interactions between different vacancy orientations induce competing tendencies:
full nematicity at dilute vacancy doping and fully polarized ferromagnetism at intermediate doping.
In the fully nematic case, all spin states are degenerate due to full permutation nonergodicity.
}

Propositions \ref{prop:evenness for WC} and \ref{prop:ergodicity for WC} together imply the following theorem that establishes the fully spin-polarized state as a unique ground state of \eqref{eq:vacancy} in the subsectors specified by conditions (a-d) of Proposition \ref{prop:ergodicity for WC}. 
Additionally, a set of gapless Fermi surface excitations is identified, establishing the metallic nature of the ground state.

\begin{theorem}
\label{thm:uniqueness for WC}
$\pmb{[}${\rm{\bf{Unique Half-Metallic Ground state from 1D Vacancies}}}$\pmb{]}$
The Hamiltonian \eqref{eq:vacancy} on an $L\times L$ triangular lattice with the periodic boundary condition (PBC) supports the fully spin-polarized ground state within each sector of the Hilbert space satisfying conditions (a-d) of Proposition \ref{prop:ergodicity for WC}.
This ground state is unique aside from $(2N+1)$-fold spin degeneracy associated with the spin rotational symmetry.
Additionally, within the fully spin-polarized sector, there exists at least one low-energy [$O(1/L)$] excitation on each $\alpha$-line, with crystal momentum $ 2k_{F}^{(\alpha)} \equiv  2 \pi m_{\alpha}/L$ relative to the ground state. 
The total number of these excitations is at least $3L$. 
\end{theorem}

\smallskip

{\it Proof of Theorem \ref{thm:uniqueness for WC}.}
As in the proof of Theorem \ref{thm:uniqueness for Lieb}, we use the Perron-Frobenius theorem for an irreducible non-positive matrix to show that the fully spin-polarized ground state established in Theorem \ref{thm:exact ferro GS for vacancy} is the unique ground state.
The non-positivity of the Hamiltonian with respect to a Perron-Frobenius basis has already been demonstrated in \eqref{eq:matrix element for vacancy}.
Furthermore, the even-permutation ergodicity within the subsector specified by conditions (a-d) of Proposition \ref{prop:ergodicity for WC} implies the spin ergodicity, ensuring the irreducibility of the Hamiltonian matrix.
Thus, the fully spin-polarized state must be the unique ground state in each specified sector of the Hilbert space.

To establish the existence of gapless excitations at `$2k_F,$' we explicitly construct them via the `adiabatic flux insertion' \cite{laughlin1981quantized, yamanaka1997nonperturbative, oshikawa2000commensurability,oshikawa2000topological} through each $\alpha$-line (a total of $3L$ lines), utilizing the $3L$ $U(1)$ subsystem symmetries (\ref{eq:subsystem symmetry 1 for WC}-\ref{eq:subsystem symmetry 3 for WC}).
We first define the `twist' operator $U_{\alpha}(k)$ on the $k$th (or $j$th) $\alpha$-line as 
\begin{align}
\label{eq:twist1}
    &U_{1}(k) \equiv \exp{\left [\frac{2\pi i}{L} \sum_{j=1}^L  j \cdot n^{\rm v}_{1}(j \v e_1 + k \v e_3)\right ]},
    \end{align}
\begin{align}
    \label{eq:twist2}
    &U_{2}(j) \equiv \exp{\left [ \frac{2\pi i}{L} \sum_{k=1}^L k \cdot n^{\rm v}_{2}(j \v e_1 + k \v e_2) \right ]},
\end{align}
    \begin{align}
    \label{eq:twist3}
    &U_{3}(j) \equiv \exp{\left [\frac{2\pi i}{L} \sum_{k=1}^L k\cdot   n^{\rm v}_{3}(j \v e_1 + k \v e_3) \right ]}.
\end{align}
These operators represent `large gauge transformations' that adiabatically insert $2\pi$ flux through the hole of each $\alpha$-line.
We then calculate the energy difference between the ground state, fully spin-polarized in $+\hat z$-direction, denoted by $\left | \Psi_0  \right >$, and the trial state $U_{\alpha}(k)\left | \Psi_0 \right >$. 
For example, for $\alpha = 1$,
\begin{align}
\label{eq:energy of twisted state}
    &\!\left < \Psi_0 \right | U_{1}(k)^{-1} H^{\rm v}_{\rm eff} U_{1}(k) \! - \! H^{\rm v}_{\rm eff}  \! \left | \Psi_0 \right >  
    \!=\!
    -t_{11} \!  \sum_{j=1}^L \! \bigg [\!  \left ( e^{2\pi i /L} \!-\! 1 \right ) 
    \nonumber \\
    &
\left <    f^{\dagger}_{(j+1) \v e_1 + k \v e_3,\uparrow}  f_{ j \v e_1 +k \v e_3, \uparrow} c^{\dagger}_{j \v e_1 +k \v e_3,1} c_{ (j+1) \v e_1 +k \v e_3,1}  \right > _{\!0} 
\!+\! {\rm c.c.} \bigg ]
\end{align}
where we used the fermionic representation of $H^{\rm v}_{\rm eff}$ \eqref{eq:fermionic vacancy}, and the expectation value $\left  <\cdots \right >_0$ is taken with respect to $\left | \Psi_0 \right >$.
Here, only terms on the  $k$th $1$-line survive since all other terms are invariant under the transformation $U_{1}(k)$.
Since  each expectation value  is a positive real number \eqref{eq:matrix element for vacancy}, their sum is also a positive real number of order $\sum_{j}\left < \cdots \right >_0 =O(L)$. 
\footnote{
More precisely, the sum is $O(N^{\rm v}_{1}(k))$; however, assuming a finite vacancy density in the thermodynamic limit, $N^{\rm v}_{1}(k)/L = O(1)$.
}
Expanding \eqref{eq:energy of twisted state} in powers of $1/L$, the first term of order $O(1)$  is purely imaginary and cancels with its complex conjugate.
Therefore, the energy difference \eqref{eq:energy of twisted state} is $O(1/L)$.

To show that the twisted state $U_{1}(k)\left | \Psi_0 \right >$ is orthogonal to $\left | \Psi_0 \right >$, we calculate its crystal momentum.
Let $T[\v e_1] = e^{-i P_1}$ be the translation operator by the lattice vector $\v e_1$.
Conjugating by $U_{1}(k), $ we obtain, $U_{1}(k)^{-1} T[\v e_1]U_{1}(k) = T[\v e_1] e^{2\pi i  N^{\rm v}_{1}(k)/L} = T[\v e_1] e^{2\pi i  m_1/L}$.
Thus,  $U_{1}(k)\left | \Psi_0 \right >$ has a crystal momentum $2\pi m_1/L \equiv 2 k_{F}^{(1)}$ relative to $\left | \Psi_0 \right >$.
This gives $L$ low-energy [$O(1/L)$] excitations ($k=1,\cdots,L$). 
Repeating the same exercise for $\alpha =2,3$ completes the proof of the existence of at least $3L$ low-energy excitations.

Note that when the vacancy filling on each line $m_{\alpha}/L$ is a rational fraction $p/q$, with $p$ and $q$ coprime, the above argument demonstrates the existence of at least $3(q-1)L$ low-energy [$O(1/L)$] excitations: $U_{\alpha}(k)\left | \Psi_0 \right >$, $U_{\alpha}(k)^2\left | \Psi_0 \right >$, $\cdots,$ $U_{\alpha}(k)^{q-1}\left | \Psi_0 \right >$. 
$\square$

\smallskip

We remark that the uniqueness of the ferromagnetic ground state can only be established within the Hilbert space subsector defined by conditions (a-d) in Proposition \ref{prop:ergodicity for WC}.
This result is more restrictive than that of Theorem \ref{thm:uniqueness for Lieb}, where the half-metallic ferromagnet is shown to be the `global' ground state for a given number of electrons.
This stronger result was possible for the solvable model \eqref{eq:solvable model} because it becomes effectively non-interacting in the fully ferromagnetic sector \eqref{eq:solvable model in ferro sector}, enabling a straightforward comparison of ground state energies across different subsectors specified by (\ref{eq:conservation_Emery1}–\ref{eq:conservation_Emery2}).
However, this approach does not apply to the vacancy Hamiltonian \eqref{eq:vacancy}.

\begin{figure}
    \centering
    \includegraphics[width=0.8\linewidth]{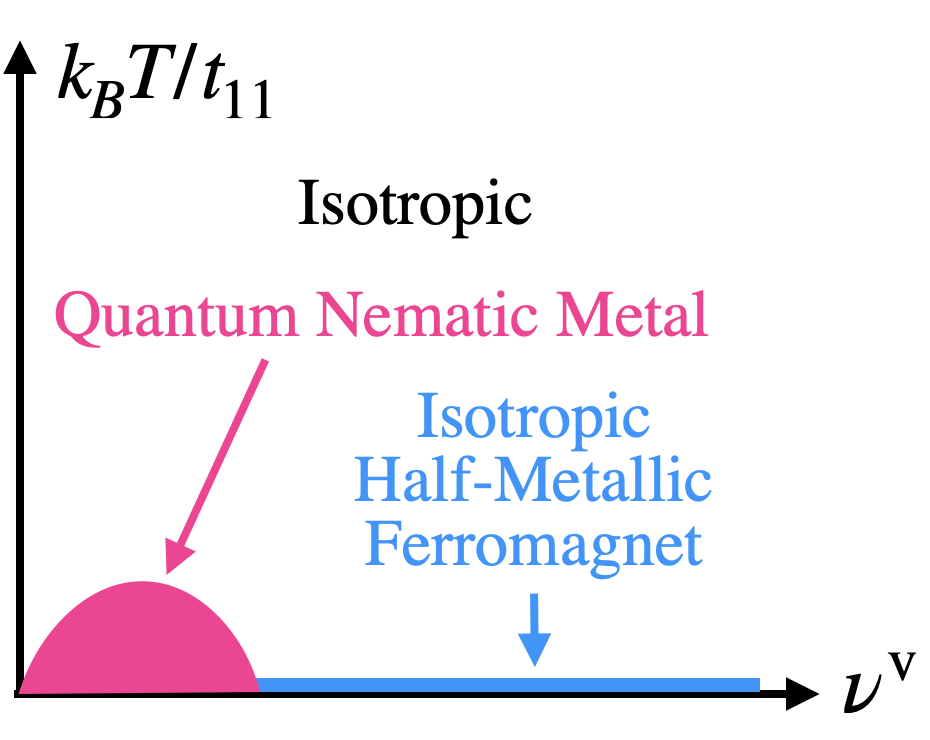}
    \caption{Schematic phase diagram of \eqref{eq:vacancy} as a function of vacancy filling $\nu^{\rm v}\equiv N^{\rm v}_{\rm tot}/L^2$ and temperature. At dilute vacancy concentrations, the quantum nematic phase is stable, which exhibits complete spin degeneracy at $T=0$. In contrast, at higher vacancy concentrations, the isotropic phase is energetically favorable,  characterized by half-metallicity.}
    \label{fig:vacancy phase diagram}
\end{figure}

An argument similar to that given around Eqs. (\ref{eq:nematic}-\ref{eq:isotropic}) suggests that in the dilute vacancy concentration ($\nu^{\rm v} < \nu^{\rm v}_{c}$), the system favors a quantum nematic phase.
At $T=0$, this phase becomes fully nematic, leading to complete spin degeneracy as in the case of the strong-coupling and semi-classical limit of the  Emery model \eqref{eq:effective Hamiltonian}.
By contrast, at higher vacancy concentrations, $\nu^{\rm v} > \nu^{\rm v}_{c}$, an isotropic phase, characterized by full spin polarization, is expected to be stable.
These considerations yield the schematic phase diagram of \eqref{eq:vacancy} as shown in Fig. \ref{fig:vacancy phase diagram}.

\section{{Exact Boson-Fermion Equivalence}}
\label{sec:Exact Boson-Fermion Correspondence}
In this section, we present an amusing corollary that follows from the evenness of ring-exchange permutations.
While a fermionic system and its corresponding hard-core bosonic counterpart typically exhibit different ground and excited state properties,
the following theorem states that they share the same eigen-spectrum when the Hamiltonian dynamics is restricted to allow only ring-exchange permutations of even parity 
\footnote{This is a formal restatement of a fact articulated by Thouless in Ref. \cite{thouless1965exchange}.}.
Our results in Secs. \ref{sec:Solvable model for a half-metal}-\ref{sec:Wigner crystal} can be understood in the context of this theorem, alongside the established fact that a system of interacting spin-$\frac 1 2$ bosons generically exhibits a fully spin-polarized ground state \cite{thouless1965exchange, eisenberg2002bosonFerromagnetism, yang2003bosonFerromagnetism},  as follows.
The 1D mobility of doped particles leads to  the `evenness of ring-exchange permutations,' 
 which in turn implies the equivalence between the fermionic systems (\ref{eq:solvable model}, \ref{eq:effective Hamiltonian} \& \ref{eq:vacancy}) and their bosonic counterparts.
 Since the bosonic models host a fully spin-polarized ground state, so too do the fermionic systems discussed in Secs. \ref{sec:Solvable model for a half-metal}-\ref{sec:Wigner crystal}.

\begin{theorem}
\label{thm:bose-fermi Correspondence}
{\rm{\textbf{$\pmb{[}$Exact Boson-Fermion Equivalence$\pmb{]}$}}} 

\noindent
Consider a number-conserving Hamiltonian $H_F \equiv H(\{f_{x,\sigma}^{\dagger},f_{x,\sigma}\})$ consisting of fermionic  operators, $f_{x,\sigma}$, where $x=1,\cdots,N_{\rm site}$ is a site index and $\sigma$ denotes  particles' internal degrees of freedom (e.g., spin). 
We assume that the Hamiltonian  dynamics does not alter each particle's internal state; accordingly, the total fermion number for each $\sigma$, given by $\sum_{x} f^{\dagger}_{x, \sigma} f_{x, \sigma}$, is conserved.
Furthermore, we require that fermionic operators in $H_F$ are `normal-ordered' as defined in the proof, which specifies a unique Boltzmannian dynamics for $f$-particles.
The corresponding Hamiltonian for hard-core bosons is defined as $H_B\equiv H(\{b_{x,\sigma}^{\dagger}, b_{x,\sigma}\})$ by replacing fermionic operators $f_{x,\sigma}$ with bosonic operators $ b_{x,\sigma}$  having hard-core  constraints $b_{x,\sigma}^{\dagger} b_{x,\sigma} \leq 1$.
{\it Then, $H_F$ and $H_B$ share the same eigen-spectrum if the Hamiltonian dynamics only allows even permutations of $f$- (or $b$-) particles.}
This boson-fermion equivalence  extends to Hamiltonians that include additional bosonic operators $a_{y}$: 
$H_F\equiv$$H(\{f_{x,\sigma}^{\dagger},f_{x,\sigma}\};\{a_{y}^{\dagger},a_{y}\}
)$ and 
$H_B\equiv H(\{b_{x,\sigma}^{\dagger}, b_{x,\sigma}\};\{a_{y}^{\dagger},a_{y}\})$.
\end{theorem}

\smallskip

{\it Proof of Theorem \ref{thm:bose-fermi Correspondence}}. 
We begin by defining the {\it normal ordering} of operators as follows.
Since the total number of $f$-particles is conserved for each $\sigma$, each term in $H_F$ contains an equal number of annihilation operators $f_{\cdot, \sigma}$ and creation operators $f_{\cdot, \sigma}^{\dagger}$ for each `spin' species $\sigma.$
The normal-ordered operator first annihilates $f$-fermions at $x_1,x_2,\cdots,x_m$ with respective spins $\sigma_1,\sigma_2,\cdots,\sigma_m$ in order, and then creates them in reverse order at positions $x_m',\cdots,x_1'$
\begin{align}
\label{eq:normal ordering}
    f^{\dagger}_{x_1', \sigma_1} f_{x_2', \sigma_2}^{\dagger} \cdots f^{\dagger}_{x_m', \sigma_m} f_{x_m, \sigma_m} \cdots f_{x_1, \sigma_1}.
\end{align}
Here, $(x_i, \sigma_i)$ and $(x_j, \sigma_j)$ [or $(x_i', \sigma_i)$ and $(x_j', \sigma_j)$]  represent distinct pairs for $i\neq j$; otherwise, the expression becomes zero due to Fermi statistics.
As noted in Footnote \cite{footnote:distinguishability}, there is an ambiguity in interpreting this term as a dynamical process for distinguishable particles.
This is because $i$th particle at $x_i$ and $j$th particle at $x_j$ can be regarded as scattering either to $x_i'$ and  $x_j'$ or to $x_j'$ and  $x_i'$, respectively.
{\it To resolve this ambiguity, we will view this term as representing a dynamical process
in which $i$th particle at $x_i$ with spin $\sigma_i$ scatters to $x_i'$.}
The Hermitian conjugate of \eqref{eq:normal ordering} must also be normal-ordered as
\begin{align}
\label{eq:Hermitian conjugate}
    f^{\dagger}_{x_1, \sigma_1} f_{x_2, \sigma_2}^{\dagger} \cdots f^{\dagger}_{x_m, \sigma_m} f_{x_m', \sigma_m} \cdots f_{x_1', \sigma_1},
\end{align}
and be interpreted as the reverse dynamical process.
Note that changing the order of the operators in the Hamiltonian results in a different description of Boltzmannian dynamics.
The theorem applies whenever we find a specific normal ordering for which the corresponding Boltzmannian dynamics induces only even permutations among $f$- (or $b$-) particles.
We assume this is the case and define 
\begin{align}
    H_F = \sum_a -t_a \cdot  h^{(a)}_F \text{ and }  H_B = \sum_a -t_a \cdot  h^{(a)}_B,
\end{align}
where $t_a$s are complex numbers, each $h_F^{(a)}$  is a normal-ordered product of $f$ and $f^{\dagger}$ as in \eqref{eq:normal ordering} or \eqref{eq:Hermitian conjugate}, and $h_B^{(a)}$ is the bosonic operator obtained by  substituting $f_{x, \sigma}\to b_{x, \sigma}$ in $h_F^{(a)}$.
It is implicitly assumed that for each $-t_a \cdot h_F^{(a)}$ in $H_F$, there is a Hermitian conjugate term $-t_a^* \cdot h_F^{(a)\dagger}$, and similarly for $H_B$.

We now proceed to prove the theorem by showing that $H_F$ and $H_B$ share the same partition function.
For simplicity, we focus on the case without additional bosonic operators $a_y$, as extending to include $a_y$ is straightforward.
We first collectively index positions and spins of $N$ Boltzmannian electrons  as $\v x = (x_i)$ and $\sigma = (\sigma_i)$, respectively.
The Fock states for the fermionic and bosonic systems are defined as 
\begin{align}
\label{eq:Fock basis 1}
    \left |\v x,\sigma  \right >_{ f} &\equiv  f_{{x_1},\sigma_1}^{\dagger}\cdots f_{{x_N},\sigma_N}^{\dagger} \vacuum,
    \\
    \label{eq:Fock basis 2}
    \left |\v x,\sigma  \right >_{ b} \, &\equiv 
     b_{{x_1},\sigma_1}^{\dagger}\cdots b_{{x_N},\sigma_N}^{\dagger} \vacuum,
\end{align}
where $\vacuum$ is a vacuum state without any particles and the pairs $(x_i,\sigma_i)$ and $(x_j,\sigma_j)$ are distinct for $i\neq j.$
The fermionic and bosonic partition functions are  written as
\begin{align}
\label{eq:partition function1}
    Z_F  &= \sum_{\v x,\sigma} {}_f\! \left <\v x,\sigma \right | e^{-\beta H_F} \left | \v x,\sigma  \right >_{ f},
    \\
    \label{eq:partition function2}
    Z_B  &= \sum_{\v x,\sigma } {}_b \! \left <\v x,\sigma  \right | e^{-\beta H_B} \left |\v x,\sigma  \right >_{b},
\end{align}
where the sum over $\v x$ is restricted to 
\begin{align}
\label{eq:I inequality}
    1 \leq x_1 \leq x_2 \leq \cdots \leq x_N \leq N_{\rm site}
\end{align} 
to avoid overcounting Fock states.
Expanding $e^{-\beta H}$ in a power series, $Z_F$ and $Z_B$ can be written as sums of terms
\begin{align}
\label{eq:taylor series1}
 &\frac{\beta^n}{n!} t_{a_1}\cdots t_{a_n} \cdot 
 {}_{f} \! \left <\v x,\sigma \right | h_F^{(a_n)}  h_F^{(a_{n-1})} \cdots  h_F^{(a_1)}\! \left |\v x,\sigma \right >_{ f},
 \\
 \label{eq:taylor series2}
 &\frac{\beta^n}{n!} t_{a_1}\cdots t_{a_n}\cdot    {}_{b} \! \left <\v x,\sigma \right | h_B^{(a_n)}  h_B^{(a_{n-1})} \cdots  h_B^{(a_1)} \left |\v x,\sigma \right >_{ b}.
\end{align}
We will now prove that $Z_F = Z_B$ by showing that each term in the expansions,
\eqref{eq:taylor series1} and \eqref{eq:taylor series2}, is identical.

To prove this claim, let us first consider the action of a normal-ordered operator 
\begin{align}
\label{eq:h_F}
h_F \equiv 
f^{\dagger}_{x_1', \sigma_1}  \cdots f^{\dagger}_{x_m', \sigma_m}  
f_{x_{i_m}, \sigma_m} \cdots f_{x_{i_1}, \sigma_1}
\end{align}
on a Fock state $\left |\v x,\sigma \right >_{f}$ \eqref{eq:Fock basis 1}.
The following identity can be straightforwardly proven by induction 
\begin{align}
\label{eq:action of normal-ordered operator}
h_F \left |\v x,\sigma \right >_{f} 
&= h_F \ 
f^\dagger_{x_1, \sigma_1}f^\dagger_{x_2, \sigma_2}
\cdots f_{x_N, \sigma_N}^\dagger 
\vacuum 
\nonumber \\
&= f^\dagger_{x_1', \sigma_1}f^\dagger_{x_2 ', \sigma_2}
\cdots f_{x_N', \sigma_N}^\dagger 
\vacuum \equiv \left | \v x', \sigma \right >_f,
\end{align}
where $\v x' \equiv (x'_i)$ is obtained by substituting each index $x_{i_k}$ in a tuple $\v x=(x_i)$ with $x_k'$.
Note that when $(x_i',\sigma_i) = (x_j',\sigma_j)$ for $i\neq j$, the expression evaluates to zero due to Fermi statistics.
An analogous identity holds for hard-core boson operators $b$.
The identity \eqref{eq:action of normal-ordered operator} describes a dynamical process of Boltzmannian particles corresponding to the normal-ordered operator \eqref{eq:h_F}
\begin{align}
\label{eq:Boltzmannian dynamics}
    (\v x, \sigma) \underset{h_F}{\rightarrow} (\v x', \sigma).
\end{align}
In this expression, we implicitly assume that $h_F$ does not annihilate the state $\left | \v x, \sigma \right >_f$.
The application of $h_F^{(a_n)}  h_F^{(a_{n-1})} \cdots  h_F^{(a_1)}$ to a state $\left |\v x,\sigma \right >_f$ then describes the following Boltzmannian dynamics
\begin{align}
    \!\!\!\! (\v x, \sigma) 
    \underset{h_F^{(1)}}{\rightarrow} (\v x^{(1)}, \sigma)
    \underset{h_F^{(2)}}{\rightarrow} (\v x^{(2)}, \sigma) 
    \underset{h_F^{(3)}}{\rightarrow}  \cdots 
    \underset{h_F^{(n)}}{\rightarrow} (\v x^{(n)}, \sigma). 
\end{align}
In order for the matrix element in \eqref{eq:taylor series1} to be nonzero, the final configuration of  Boltzmannian particles, $(\v x^{(n)}, \sigma)$, must be some permutation of their initial configuration
\begin{align}
    x_{\pi(i)}^{(n)} =x_i \ \text{ and } \ \sigma_{\pi(i)}  = \sigma_i \text{ for } i=1,\cdots,N, 
\end{align}
for some $\pi \in {\rm S}_N$.
In this case, the matrix element becomes
\begin{align}
\label{eq:fermion overlap}
 \!\!\!\! {}_{ f}\! \left <\v x,\sigma \right |
 h_F^{(a_n)} \! \cdots h_F^{(a_1)}\! \left |\v x,\sigma \right >_{ f} 
 \!=\! {}_{ f}\! \left <\v x,\sigma  | \v x_n,\sigma \right >_{ f} 
 \!=\! {\rm sgn}(\pi).
\end{align}
In contrast, the same consideration for hard-core bosons leads to 
\begin{align}
\label{eq:boson overlap}
  {}_{ b}\! \left <\v x,\sigma \right |
 h_B^{(a_n)} \! \cdots h_B^{(a_1)}\! \left |\v x,\sigma \right >_{b} 
 \!=\! {}_{b}\! \left <\v x,\sigma  | \v x_n,\sigma \right >_{b} 
  = 1.
\end{align}
It follows immediately that if only even permutations are allowed by the Hamiltonian dynamics, then ${\rm sgn}(\pi)=1 $, and each term in the fermionic and bosonic partition function, \eqref{eq:taylor series1} and \eqref{eq:taylor series2}, are identical.
This completes the proof of the theorem.
$\square$

A direct corollary of this theorem and Proposition \ref{prop:evenness for Lieb}  is that the bosonic version of the solvable Hamiltonian \eqref{eq:solvable model} shares exactly the same spectrum as its fermionic counterpart, and thus hosts fully spin-polarized Fermi gas ground state.
Since the underlying particles are bosons, we refer to this state as a (ferromagnetic) {\it Bose Fermi gas.}
In particular, this Bose Fermi gas features straight-line Fermi surfaces [Fig. \ref{fig:fermi surface}].
By introducing additional interactions, such as next-nearest neighbor repulsion \eqref{eq:NNN interactions}, interacting Fermi phases (such as a {\it Bose Fermi liquid} phase) can also be realized in a system consisting of bosonic particles.
Similarly, the bosonic version of 1D vacancy Hamiltonian \eqref{eq:vacancy} is expected to host a (ferromagnetic) Bose Fermi liquid phase when the vacancy concentration is not dilute.

\section{Conclusion}
\label{sec:discussion}
We have proposed a universal kinetic mechanism for half-metallic ferromagnetism, arising from the interplay between strong interactions that prohibit double occupancy and the 1D mobility of doped particles.
We illustrate this mechanism concretely in two representative systems with 1D particles: the solvable model on a Lieb lattice \eqref{eq:solvable model} [Sec. \ref{sec:Solvable model for a half-metal}] and the 1D vacancy Hamiltonian in a Wigner crystal \eqref{eq:vacancy} [Sec. \ref{sec:Wigner crystal}].
(The former can be viewed as a solvable deformation of the strong-coupling, semi-classical limit of the Emery model \eqref{eq:effective Hamiltonian} discussed in Sec. \ref{sec:Emery model}.)
Our results show that 1D mobility and strong interactions generically constrain the Hamiltonian dynamics to induce only even permutations among electrons (Propositions \ref{prop:evenness for Lieb} \& \ref{prop:evenness for WC}).
Combined with the Thouless rule---which states that kinetic processes involving even (odd) permutations mediate ferromagnetism (antiferromagnetism)---this implies the existence of a ferromagnetic ground state (Theorems \ref{thm:exact ferro ground state} \& \ref{thm:exact ferro GS for vacancy}). 
The uniqueness of the ferromagnetic state, demonstrated in Theorems \ref{thm:uniqueness for Lieb} \& \ref{thm:uniqueness for WC}, results from spin ergodicity within the Hilbert space subsector specified by $S^z_{\rm tot}$ and subsystem symmetries [(\ref{eq:conservation_Emery1}-\ref{eq:conservation_Emery2}) or (\ref{eq:subsystem symmetry 1 for WC}-\ref{eq:subsystem symmetry 3 for WC})].
In the first-quantization formalism, this spin ergodicity, in turn, arises from the even-permutation ergodicity of the underlying Hamiltonian dynamics (Propositions \ref{prop:permutation ergodicity OBC}, \ref{prop:permutation ergodicity PBC} \& \ref{prop:ergodicity for WC}).
Generalization of the above results to  $SU(N)$ spins [instead of $SU(2)$] is straightforward under this framework.
The metallic nature of the ferromagnetic state follows from the standard Lieb-Schultz-Mattis-type argument (Theorem \ref{thm:uniqueness for WC}).
For the vacancy Hamiltonian \eqref{eq:vacancy}
and the strong-coupling, semi-classical limit of the Emery model \eqref{eq:effective Hamiltonian}, we discuss the competing tendency toward full nematicity with complete spin degeneracy in the dilute doping regime (Fig. \ref{fig:Emery phase diagram} \& \ref{fig:vacancy phase diagram}).
Finally, as a corollary, we demonstrate that 1D mobility and the resulting evenness of ring exchange permutations provide a general condition for the exact boson-fermion equivalence.

While we have demonstrated these results in a few representative systems [\eqref{eq:solvable model}, \eqref{eq:effective Hamiltonian} \& \eqref{eq:vacancy}] for concreteness, our proposed mechanism for itinerant ferromagnetism is universal and does not rely on specific geometry or dimensionality.
This work thus establishes a thermodynamically robust kinetic mechanism  for half-metallic ferromagnetism in electronic models with spin-independent Coulomb interactions. 
\footnote{
Ferromagnetic ground states in flat-band and related models typically exhibit a charge gap and therefore do not constitute metallic ferromagnetism. 
In such models, ferromagnetism arises as a weak-coupling phenomenon, occurring for any small $U>0$ due to singular band structures.
See Refs. \cite{mielke1991ferromagnetic, mielke1993ferromagnetism, tasaki1998nagaoka, tasaki2003ferromagnetism, tasaki2020physics}, for examples.
Lieb's ferrimagnetism on bipartite lattices \cite{lieb1989two} or quantum Hall ferromagnetism \cite{sondhi1993skyrmions} also describe an insulating ferromagnetism resulting from quenched kinetic energy and occurs for any weak interactions $U>0$.
These forms of ferromagnetism can be broadly categorized as resulting from a generalized Hund's mechanism.
}
\footnote{Although Li et al. \cite{li2014exact} established the existence of a half-metallic ground state in the thermodynamic limit, their ferromagnetism is driven fundamentally by on-site ferromagnetic Hund's coupling rather than by spin-independent Coulomb interactions. 
In their model, all spin states become degenerate in the absence of Hund's coupling.
On a separate note, a generalized Hubbard model for a half-metallic ground state is proposed in Ref. \cite{tanaka2007metallic} (see also related discussions in \cite{tasaki2020physics}).
However, the proposed model is rather complicated, and the ``physical'' mechanism behind the half-metallic ferromagnetism in their model is unclear.
}
Whether any experimental system exhibits half-metallic ferromagnetism driven by our proposed mechanism is an intriguing open question worth exploring in the future.

\section*{Acknowledgments}
K-S.K. is grateful to Steven A. Kivelson for numerous insightful discussions during the development of this work and for directing him to Ref. \cite{kivelson2004quasi1D}.
K-S.K. also thanks helpful discussions with Zhaoyu Han, Giuseppe de Tomasi, Pavel Nosov, Eduardo Fradkin, Rafael Fernandes and Erez Berg, and the Massachusetts Institute of Technology for its hospitality where this work was initiated.
We also appreciate Hosho Katsura and Matthew O'Brien for uesful comments on the draft.
K-S.K. was supported, in part, by NSF-BSF award DMR2310312 at Stanford University, and by the Anthony J. Leggett Postdoctoral Fellowship at the University of Illinois  Urbana-Champaign.

\appendix

\section{Proof of Propositions \ref{prop:permutation ergodicity OBC} and \ref{prop:permutation ergodicity PBC}
}
\label{app:proof of ergodicity for Lieb}

In this section, we provide the proof of the even-permutation ergodicity of Proposition \ref{prop:permutation ergodicity OBC}.
For Proposition \ref{prop:permutation ergodicity PBC}, we only  outline the proof as it is essentially the same as the proof for Proposition \ref{prop:permutation ergodicity OBC}.
For better readability, we repeat the propositions here.

\smallskip

\noindent
{\rm\textbf{Proposition II.3.} 
{\rm\textbf{$\pmb{[}$Even-Permutation Ergodicity; OBC$\pmb{]}$}} 
Consider the Hamiltonian \eqref{eq:solvable model} on an $L_x \times L_y$ Lieb lattice under the open boundary condition (OBC), with dangling bonds at the boundary [see {Fig. \ref{fig:solvable Lieb lattice} (a)}]. 
Consequently, there are $L_x+1$ and $L_y+1$ bond $p$ orbitals on each row and column, respectively. 
For electronic configurations that satisfy the following condition,
\begin{itemize}
    \item[(a)] $0<X(y)<L_x$ and $0<Y(x)<L_y $ for each $y$ and $x$, respectively, meaning that each row and column contains at least one electron and two vacancies,
\end{itemize}
{\it any even permutation (and none of the odd permutations) of $N$ Boltzmannian electrons can be induced} by repeated applications of terms in \eqref{eq:solvable model}.

\smallskip

{\it Proof of Proposition \ref{prop:permutation ergodicity OBC}. }
We will use elementary 3-cycles introduced in Fig. \ref{fig:solvable Lieb lattice} (b) as building blocks for generating the alternating group ${\rm A}_N$.
Proving whether or not a set of cycles generate the entire alternating group ${\rm A}_N$ can be rather cumbersome.
We will use the following result (Lemma 2 in Ref. \cite{wilson1974graph}) to simplify the proof.

\smallskip

\begin{lemma}
\label{lemma:transitivity}
    Let $\Sigma$ be a set of 3-cycles on a finite set $X$,  where $X$ has $N\geq 3$ elements, and let $\left <\Sigma \right >$ denote the subgroup of the symmetric group ${\rm S}_X$ generated by $\Sigma$. Then the following are equivalent:
\begin{align}
    \label{eq:alternating group1}
    {\rm (i)}\ &\left < \Sigma \right > = {\rm A}_X. \\
    \label{eq:alternating group2}
    {\rm (ii)}\ &\left < \Sigma \right >\   \textit{is transitive on } X. 
\end{align}
Here, ${\rm A}_X$ denotes an alternating group on $X$. 
The group action $G$ on $X$ is said to be {\it transitive} if 
\begin{align}
\label{eq:transitivity}
   \textit{for any $\alpha, \beta \in X$, there exists $g \in G$  such that }  g \cdot \alpha = \beta. 
\end{align}
\end{lemma}

We refer the interested readers to Ref. \cite{wilson1974graph} for the proof of the above statement.
We will use some set of elementary 3-cycles on Boltzmannian electrons's index set $X = \{1,2,...,N \}$ as a generating set $\Sigma$ and show that $\left < \Sigma \right >$ is transitive on $X$.
This is enough to show that any even permutation can be induced by the dynamics of the Hamiltonian for the configurations satisfying condition (a) of the proposition.
Combined with Proposition \ref{prop:evenness for Lieb}, which states that no odd permutations can be induced, the even-permutation ergodicity follows.

It is convenient to prove the transitivity for  electron-vacancy configurations satisfying condition (a) in which all the dangling edges are occupied by vacancies.
This is always possible when condition (a) is satisfied  since the positional ergodicity holds for model \eqref{eq:solvable model}.
We will refer to such a configuration as a {\it canonical charge configuration for the OBC}.

\begin{figure}
    \centering
    \includegraphics[width= \linewidth]{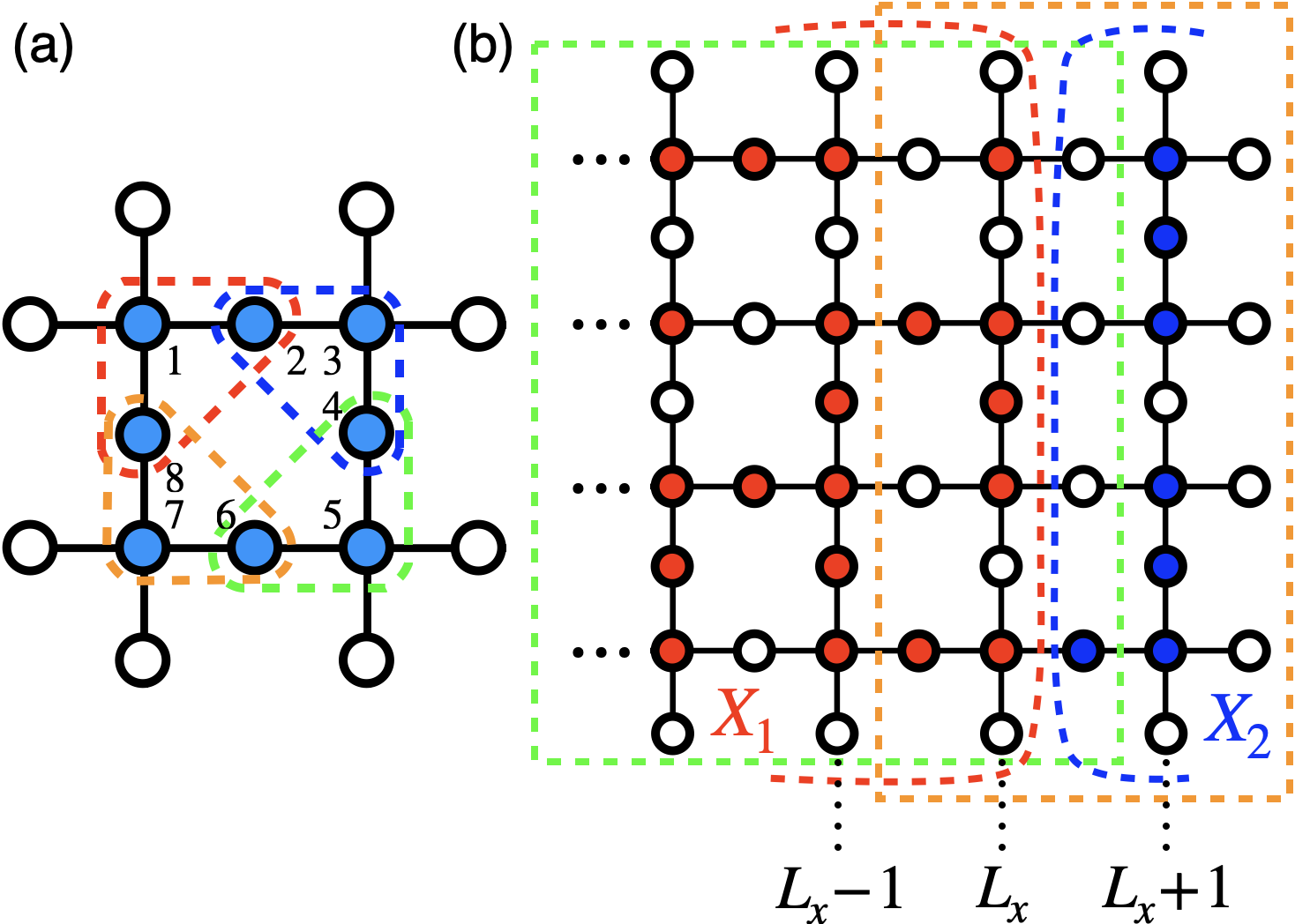}
    \caption{(a) A canonical charge configuration in $2\times 2$ geometry used as the base case of the induction proof of even-permutation ergodicity in Proposition \ref{prop:permutation ergodicity OBC}.
    Four elementary $3$-cycles---$( 1\ 2\ 8),\ (2\ 3\ 4),\ (4\ 5\ 6)$ and $(6\ 7\ 8)$---highlighted in red, blue, green and orange dashed lines, respectively, generate the alternating group on the $8$ numbered electrons.
    (b) A canonical charge configuration in $(L_x+1)\times L_y$ geometry used as the inductive step of the proof. 
    Even-permutation ergodicity is demonstrated for the electrons' index set $X$ on this geometry, assuming the even-permutation ergodicity of subsets of $X$ on smaller blocks $L_x \times L_y$ (green dashed box) and $2 \times L_y$ (orange dashed box).
    In doing so, we decompose $X=X_1 \sqcup X_2$ into two disjoint sets $X_1$ (red sites) and $X_2$ (blue sites). White sites are empty sites.
    }
    \label{fig:induction Emery}
\end{figure}

The proof proceeds by induction.

{\it Base case:}
The base case is for a $2 \times 2$ Lieb lattice with $4$ vertex sites and $12$ bond sites. 
The only possible filling that satisfies condition (a) is when $X(y) = Y(x) = 1$ for $x,y = 1,2$, with a total of $N= 8$ electrons.
It is straightforward to verify that the set of four 3-cycles depicted in Fig. \ref{fig:induction Emery} (a) generates a transitive group action on the electrons' index set.

{\it Induction step:}
Let us now assume that for canonical charge configurations on $L_x \times L_y$ and smaller Lieb lattices, some set of elementary 3-cycles generates a transitive group action on the electrons' index set.
We would like to show that the same statement holds for canonical charge  configurations on $L_x \times (L_y+1)$ and $(L_x +1) \times L_y$ Lieb lattices.
Let us consider the latter case, $(L_x +1 )\times L_y$, without loss of generality.
We will divide the electronic index set into two disjoint sets $X=X_1 \sqcup X_2$: one set ($X_1$) contains all the electrons on and left of the $x=L_x$ column and the other set ($X_2 \equiv X\setminus X_1$) contains all the electrons right to the $x = L_x$ column. 
In addition, we require that there is at least one $p_x$-electron on each row to the left of the  $x= L_x$ column. 
See Fig. \ref{fig:induction Emery} (b) for illustration of such a configuration.
We need to consider three cases to prove the transitivity of the permutation group \eqref{eq:transitivity}: (1) $\alpha,\ \beta \in X_1$; (2) $\alpha,\ \beta \in X_2$;
and (3) $\alpha \in X_1$ and $\beta \in X_2$.

In order to show the transitivity for case (1), we construct a canonical charge configuration that places all electrons in $X_1$, and no others, on a  smaller $L_x \times L_y$ block, with dangling bonds at the boundary.
This is achieved by moving the leftmost vacancy  among those in $x\geq L_x +\frac 1 2 $ to  $x=L_x + \frac 1 2$ for each row.
By inductive hypothesis, the transitivity for case (1) is automatically satisfied.
Case (2) can be analogously proven by taking the rightmost $2 \times L_y$ block, with the dangling bonds at its boundary, and moving some electrons in $X_1$ such that the charge configuration in the $2 \times L_y$ block is made canonical. This can be done by first filling an electron on each bond in $x=L_x + \frac 1 2 $ column (if it is not filled already) and then by placing a vacancy on each bond in $x= L_x - \frac 1 2 $ by applying terms in the Hamiltonian \eqref{eq:solvable model}, without moving any vacancy on the top, bottom and right boundaries.
This procedure does not change the locations of electrons in $X_2$.
The transitivity for case (2) then follows from the inductive hypothesis on the $2 \times L_y$ block.
For case (3), $\alpha \in X_1$ and $\beta \in X_2$, we use a combination of the procedures used in cases (1) and (2).
First, by applying a sequence of elementary 3-cycles, we move the $\alpha$ electron to the site $(L_x,1)$ which is possible to the transitivity proven for case (1). 
Then, electron on the site $(L_x,1)$ can be moved to any location in the rightmost $2 \times L_y$ block by the procedure described for case (2).
This completes the induction proof that the group action generated by the elementary 3-cycles is transitive on the electrons' index set $X$ \eqref{eq:alternating group2}.

By Lemma \ref{lemma:transitivity}, the transitivity of the group action in turn implies that the entire alternating group can be generated.
That none of the odd permutations can be generated follows from Proposition \ref{prop:evenness for Lieb}.
$\square$
\smallskip

\noindent
{\rm\textbf{Proposition II.4.}}
$\pmb{[}${\rm\textbf{Even-Permutation Ergodicity; PBC}}$\pmb{]}$ 
Consider the Hamiltonian \eqref{eq:solvable model} with the periodic boundary condition (PBC) on an $L_x \times L_y$ Lieb lattice.
We impose the same condition as in Proposition \ref{prop:evenness for Lieb}:
\begin{itemize}
    \item[(a)] The number of electrons on each $y$th row and $x$th column, $L_x + X(y)$ and $L_y + Y(x)$, respectively, must be odd.
\end{itemize}
Additionally, if the number of electrons occupying $p$ orbitals satisfy the following condition:
\begin{itemize}
    \item[(b)]  $0<X(y)<L_x$ and $0<Y(x)<L_y $ for each $y$ and $x$, respectively, meaning that each row and column contains at least one electron and one vacancy,
\end{itemize}
then the Hamiltonian dynamics is even-permutation-ergodic.

\smallskip
{\it Proof of Proposition \ref{prop:permutation ergodicity PBC}.}
The proof for the PBC is analogous to that for the OBC, with a few modifications.
First, a {\it canonical charge configuration for the PBC} is defined as the configuration that satisfies condition (b) and where all the bonds on the column $x=\frac 1 2$ and row $y = \frac 1 2 $ are occupied by vacancies and.
A set of elementary 3-cycles [Fig. \ref{fig:solvable Lieb lattice} (b)] is similarly defined. 
The transitivity of the permutation group on electrons' index set can be demonstrated by induction in an analogous manner. 
The details are omitted here.
$\square$

\section{Proof of Proposition \ref{prop:ergodicity for WC} }
\label{app:ergodicity for WC} 

\begin{figure*}[t]
    \centering
    \includegraphics[width= \textwidth]{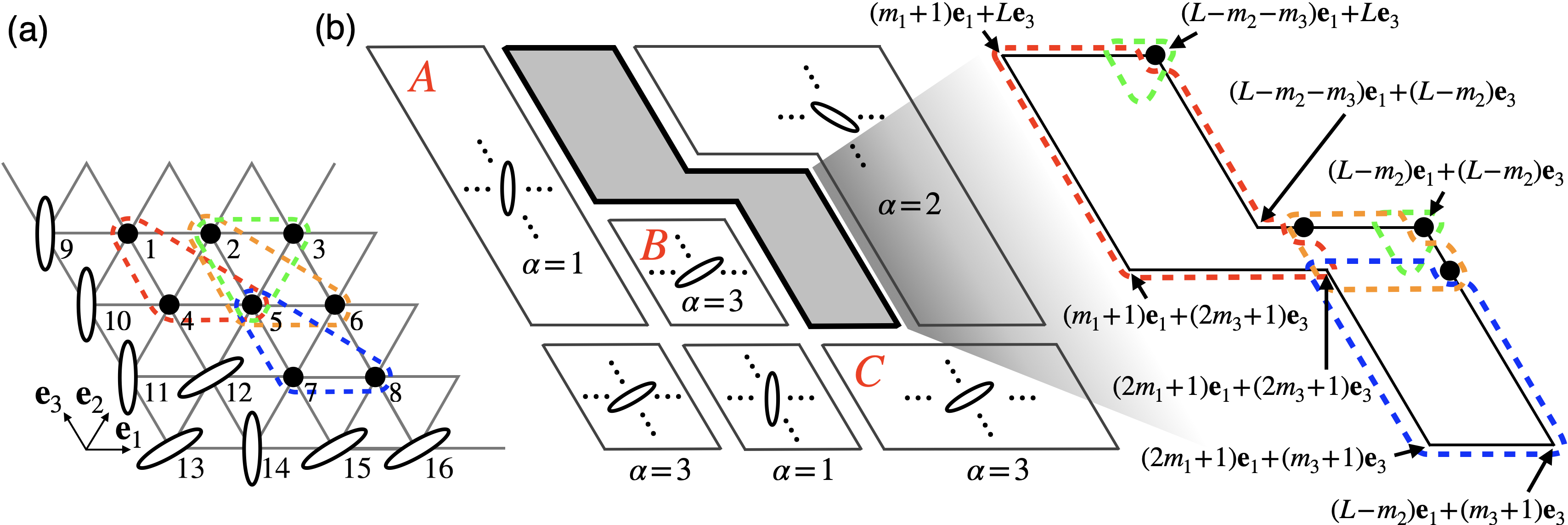}
    \caption{(a) A canonical vacancy configuration for the $4\times 4$ geometry with $ m_3 = m_1 = 1$ and $m_2 =0$, the smallest system that satisfies conditions (a-d) of Proposition \ref{prop:ergodicity for WC}.
    Four elementary $3$-cycles---$(1\ 5\ 4)$, $(2\ 6\ 5)$, $(2\ 3\ 5)$ and $(5\ 8\ 7)$, highlighted in red, orange, green and blue dashed lines, respectively---generate the alternating group of 8 numbered electrons.
    (b) A schematic of a canonical vacancy configuration for $L\times L$ system, where vacancies are positioned as specified by (\ref{eq:alpha1-1}-\ref{eq:alpha3-3}). 
    Each white block is filled with vacancies with a specified orientation $\alpha$, while the gray block is filled with electrons.
    The inset shows an expanded view of the gray area, with the coordinates specified for the corner locations.
    Conditions (c-d) of Proposition \ref{prop:ergodicity for WC} ensure that each outer edge of the gray area contains more than two sites, allowing room for the elementary 3-cycle illustrated in Fig. \ref{fig:vacancy exchange}.
    By making correlated movements of the vacancies contained in blocks $A$, $B$ and $C,$ one can induce overlapping sets of 3-cycles [red, orange, green and blue dashed lines] that cover all electrons. 
    Certain electrons indicated by black dots in the inset figure indicate that they are contained in certain types of 3-cycles but not in others. 
    For example, the top black dot is not contained within the red dashed line but is contained within the green 3-cycle.
    }
    \label{fig:proofVacancy}
\end{figure*}

In this section, we provide the proof of the even-permutation ergodicity of Proposition \ref{prop:ergodicity for WC}, which we restate below for convenience.
\smallskip

\noindent
{\rm {\textbf{Proposition IV.3.}}}
{\rm{\bf{$\pmb{[}$Even-Permutation Ergodicity$\pmb{]}$}}}
Consider the Hamiltonian \eqref{eq:vacancy} on an $L \times L$ triangular lattice with the periodic boundary condition (PBC). 
We require that 
\begin{itemize}
    \item[(a)] the number of $\alpha$-vacancies on each $\alpha$-line is the same: $ N^{\rm v}_{\alpha}(i) \equiv m_{\alpha}$ for each $i$;
    \item[(b)] $L\! -m_{\alpha}$ is odd unless $m_{\alpha }\! =\!0$ [as in Proposition \ref{prop:evenness for WC}];
    \item[(c)] there exist at least two 
    vacancy orientations; 
    if only two orientations are present, we also require that $2m_{\alpha} \leq L-  2$ for each $\alpha$;
    and
    \item[(d)] $\sum_{\alpha=1}^3 m_{\alpha} \equiv m \leq L-2$.
\end{itemize}
Then, the even-permutation ergodicity is satisfied under the 1D vacancy dynamics \eqref{eq:vacancy} within the sector connected to the {\it canonical vacancy configuration} as defined in (\ref{eq:alpha1-1}-\ref{eq:alpha3-3}) [see also Fig. \ref{fig:JW transformation}].

\smallskip

{\it Proof of Proposition \ref{prop:ergodicity for WC}.}
The proof of this proposition is analogous to that of Proposition \ref{prop:permutation ergodicity OBC} given in Appendix \ref{app:proof of ergodicity for Lieb}.
We will show that a certain set of elementary 3-cycles generates a transitive group action on the electrons' index set.
By Lemma \ref{lemma:transitivity}, this implies that the generated group is the alternating group on the electrons' index set.
Throughout the proof we assume, without loss of generality, $m_3 \geq m_1 \geq 1$ if only two orientations present and $m_2 \geq m_3 \geq m_1 \geq 1$ if all three orientations are present.

The smallest system satisfying conditions (a-d) is a $4\times 4$ triangular lattice with $m_1 = m_3 =1$ and $m_2 = 0$.
We begin by identifying the set of four $3$-cycles---$(1\ 5\ 4)$, $(2\ 6\ 5)$, $(2\ 3\ 5)$ and $(5\ 8\ 7)$---highlighted in red, orange, green and blue dashed lines in Fig. \ref{fig:proofVacancy} (a).
The red 3-cycle, $(1\ 5\ 4)$, is obtained through the correlated motions of vacancies 10 and 12, similar to the process illustrated in Fig. \ref{fig:vacancy exchange}.
The orange 3-cycle, $(2\ 6\ 5)$, results from  similar movements of  vacancies 10 and 15.
To obtain the blue 3-cycle, $(5\ 8\ 7)$, correlated motions of  vacancies 11 and 15 are required, but vacancy 12 is blocking the movement of  vacancy 11.
To resolve this, we first move  vacancy 12 out of the way, say to  site 4, then perform the correlated movements of  vacancies 11 and 15, and finally move vacancy 12 back to its original position.
Finally, the green 3-cycle is obtained as follows:
first, move vacancy 16 to the position of electron 3, displacing electron 3 to the position of electron 6; next, make correlated movements of vacancies 10 and 15 to perform the elementary 3-cycle; finally, return vacancy 16  to its initial location.

It is straightforward to see that the set of four $3$-cycles obtained through the above procedures generates a transitive group action on the electrons' index set, and thus the alternating group on this set by Lemma \ref{lemma:transitivity}.
Additionally, no odd permutations can be induced due to Proposition \ref{prop:evenness for WC}, establishing the even-permutation ergodicity for this smallest geometry.

The proof for larger geometries is analogous but slightly more complicated.
We first arrange the vacancies in a canonical configuration as shown in Fig \ref{fig:proofVacancy} (b), with all electrons contained within the gray area (see the inset for the expanded view).
Note that conditions (c-d) of Proposition \ref{prop:evenness for WC} are important to ensure that every outer edge of the gray area contains more than two sites, which, in turn, allows room for the elementary 3-cycle through the procedure illustrated in Fig. \ref{fig:vacancy exchange}.
We would like to show that the alternating group of electrons' index set can be generated from a certain set of elementary 3-cycles.
First, by performing correlated movements of vacancies contained in blocks $A$ and $B$, one can induce an overlapping set of 3-cycles of electrons contained in the red dashed line of the figure inset.
This procedure is analogous to the one performed to induce the red 3-cycle in Fig. \ref{fig:proofVacancy} (a).
Such an overlapping set of 3-cycles induces a transitive group action on the index set of electrons contained within the red dashed line of Fig \ref{fig:proofVacancy} (b).
Similarly, the transitive group action on the electrons contained in orange and blue dashed lines can be generated by the correlated movements of vacancies in blocks $A$, $B$ and $C$, through the analogous procedures performed for orange and blue 3-cycles in Fig. \ref{fig:proofVacancy} (a), respectively.
Finally, there are two electrons located at $(L-m_2-m_3)\v e_1 + L \v e_3$ and $(L-m_2)\v e_1 + (L-m_2) \v e_3$  [one if only two vacancy orientations are present] not covered by red, orange and blue areas.
3-cycles involving these electrons can be induced by the analogous procedure performed for the green 3-cycle in Fig. \ref{fig:proofVacancy} (a).
It is straightforward to see that the set of all these 3-cycles generate the transitive group action and thus the alternating group on the total electrons' index set.
Since no odd permutations of these electrons can be induced, the even-permutation ergodicity  is established for configurations satisfying conditions (a-d).
This completes the proof of Proposition \ref{prop:ergodicity for WC}.
$\square$

\bibliography{ref.bib}

\begin{thebibliography}{74}
\bibitem [{foo({\natexlab{a}})}]{footnote:filling}%
  \BibitemOpen
  \bibinfo {note} {This half-filling constraint can be relaxed to allow some vacant $d$ orbitals without much affecting the conclusions of Proposition \ref{prop:evenness for Lieb} and Theorem \ref{thm:exact ferro ground state}. [As noted later in Footnote \cite{footnote:Invariant}, the fermion parity condition (a) of Proposition \ref{prop:evenness for Lieb} can also be relaxed.] Consequently, the fully spin-polarized state remains one of the ground states of \eqref{eq:solvable model} even in the presence of some vacant $d$ orbitals. However, the conditions for the uniqueness of the ferromagnetic state, as stated in Propositions \ref{prop:permutation ergodicity OBC}-\ref{prop:permutation ergodicity PBC} and Theorem \ref{thm:uniqueness for Lieb}, require modification when some $d$ orbitals are unoccupied. For simplicity, we restrict our analysis in this paper to the case where every $d$ orbital is singly occupied.}\BibitemShut {Stop}%
\bibitem [{foo({\natexlab{c}})}]{footnote:distinguishability}%
  \BibitemOpen
  \bibinfo {note} {Interpreting a Hamiltonian for identical particles as one for distinguishable particles may seem ambiguous, particularly when the terms in the Hamiltonian involve more than two particles as in \eqref{eq:solvable model}. This ambiguity can be resolved by assigning each `normal-ordered' operator a unique dynamical process involving Boltzmannian particles. The precise definition of a normal-ordered operator and its corresponding assignment of Boltzmannian dynamics are deferred to Sec. \ref{sec:Exact Boson-Fermion Correspondence} around \eqref{eq:normal ordering}. For now, we simply interpret the Hamiltonian \eqref{eq:solvable model} as describing the dynamical process in which Boltzmannian electron with spin $\sigma$ occupying the $p$ orbital at site $\v R- {\hat \delta}/{2}$ moves to the $d$ orbital at $\v R$, while another electron with spin $\sigma'$ occupying the $d$ orbital at $\v R$ moves to the $p$ orbital at $\v R +{\hat \delta}/2$.}\BibitemShut {Stop}%
\bibitem [{foo({\natexlab{d}})}]{footnote:Invariant}%
  \BibitemOpen
  \bibinfo {note} {As noted in Footnote \cite{footnote:filling}, relaxing the half-filling condition for $d$ orbitals does not affect the conclusion for the OBC since the argument regarding the `invariant' \eqref{eq:invariant for Lieb} remains valid. For the PBC, the fermion parity condition (a) can be further relaxed for rows and columns that contain a vacant $d$ site. This is because the cyclic permutations involving an even number of vacancies, which would spoil the invariant, cannot occur: the vacant $d$ site acts as a roadblock. For further details, see the argument around \eqref{eq:cyclic permutation of vacancies} for more detail.}\BibitemShut {Stop}%
\bibitem [{foo({\natexlab{e}})}]{footnote:BCforLieb}%
  \BibitemOpen
  \bibinfo {note} {There are two subtleties in the proof for the periodic boundary condition (PBC). (1) The first involves the gauge transformation \eqref{eq:gauge transformation for solvable model}, which introduces a phase factor $(-1)^{L_x}$ [or $(-1)^{L_y}$] to the hopping terms across the periodic boundary in \eqref{eq:solvable'}. (2) The second involves the reordering of $p$-operators when evaluating \eqref{eq:matrix element for $T$}, which introduces additional phase factor $(-1)^{X(y)-1}$ [or $(-1)^{Y(x)-1}$] for matrix elements that move a $p$-electron across the periodic boundary. Since the fermion parity on each row and column is assumed to be odd, the combined phase factor is $(-1)^{L_x+X(y)-1} =1$ [or $(-1)^{L_y+Y(x)-1} =1$]. This ensures that all non-zero Hamiltonian matrix elements have a uniformly negative sign.}\BibitemShut {Stop}%
\end{thebibliography}

\end{document}